\documentclass[prb, twocolumn, showpacs, amsmath, amssymb]{revtex4-1}

\usepackage{amssymb,amsfonts,amsmath} 
\usepackage{graphicx}
\usepackage{color}
\usepackage{hyperref}
\usepackage{longtable}
\usepackage{bbm}
\begin{document} 

\title{The spectral function of the Tomonaga-Luttinger model revisited: \\ power laws and universality}

\author{L.\ Markhof}  
\affiliation{Institut f{\"u}r Theorie der Statistischen Physik, RWTH Aachen University 
and JARA---Fundamentals of Future Information
Technology, 52056 Aachen, Germany}

\author{V.\ Meden} 
\affiliation{Institut f{\"u}r Theorie der Statistischen Physik, RWTH Aachen University 
and JARA---Fundamentals of Future Information
Technology, 52056 Aachen, Germany}

\begin{abstract} 
We reinvestigate the momentum-resolved single-particle spectral function of the Tomonaga-Luttinger model. 
In particular, we focus on the role of the momentum-dependence of the two-particle interaction $V(q)$. 
Usually, $V(q)$ is assumed to be a constant and integrals are regularized in the ultraviolet 
`by hand' employing an ad hoc procedure. As the momentum dependence of the interaction is irrelevant in 
the renormalization group sense this does not affect the universal low-energy properties of the model, 
e.g. exponents of power laws, if {\it all} energy scales are sent to zero. 
If, however, the momentum $k$ is fixed away from the Fermi momentum $k_{\rm F}$, with 
$|k-k_{\rm F}|$ setting a nonvanishing energy scale, the details of $V(q)$ start to matter. 
We provide strong evidence that any curvature of the two-particle interaction at small 
transferred momentum $q$ destroys power-law scaling of the momentum resolved spectral function 
as a function of energy. Even for $|k-k_{\rm F}|$ much smaller than the momentum space range 
of the interaction the spectral line shape depends on the details of $V(q)$. The significance 
of our results for universality in the Luttinger liquid sense, for experiments on quasi 
one-dimensional metals, and for recent results on the spectral function of 
one-dimensional correlated systems taking effects of the curvature of the single-particle 
dispersion into account (`nonlinear LL phenomenology') is discussed.  
\end{abstract}

\date{\today} 
\maketitle

\section{Introduction}

\subsection{Luttinger liquid universality and the Tomonaga-Luttinger model}

It is well established that the Tomonaga-Luttinger model (TLM)\cite{Tomonaga50,Luttinger63} 
with linear single-particle dispersion and a two-particle interaction potential 
$V(q)$ which is finite at zero momentum 
transfer $q=0$ forms the infrared fixed point under renormalization group (RG) flow of a 
large class of gapless one-dimensional (1d) models of correlated fermions.\cite{Solyom79} 
This is the essence of the much celebrated Luttinger liquid (LL) 
universality.\cite{Haldane81,Voit95,Giamarchi03,Schoenhammer05}  
It implies that the low-temperature thermodynamic properties as well as the 
low-energy spectral functions of a model belonging to the LL universality class are 
equivalent to the ones of the TLM. Understanding the low-energy physics of the latter is 
thus of crucial importance. Fortunately, using bosonization\cite{Haldane81,vonDelft98,Schoenhammer05} 
it is possible to derive exact and closed analytical expressions for thermodynamic 
observables such as the specific heat or the compressibility as well as for space-time 
correlation functions of the TLM. From the latter spectral functions can be computed by 
Fourier transform. 

The bosonization expressions for correlation functions of the TLM depending 
on position $x$ and time $t$ generically 
contain integrals over momenta. Within constructive bosonization, which is based on operator 
identities,\cite{Haldane81,Schoenhammer05} these are naturally regularized in the 
ultraviolet by the momentum-space range $q_{\rm c}>0$  of the two-particle potential $V(q)$. 
For time-dependent correlation functions 
the momentum integrals cannot be performed even if a specific form of $V(q)$ is assumed.\cite{Meden99} 
However, one can show that the momentum dependence of the interaction is RG irrelevant.\cite{Solyom79} 
This is employed to justify the following procedure: In the final expressions for the  space-time correlation 
functions $V(q)$ is routinely replaced by a constant. As a consequence the momentum integrals become 
divergent in the ultraviolet. These divergences are regularized (`by hand') applying an ad hoc 
procedure.\cite{Luther74,Voit95,Giamarchi03} We already now emphasize that this regularization is not unique.
After these steps the momentum integrals can be performed 
and integral-free expressions for space-time correlation functions are obtained. In field-theory 
inspired phenomenological bosonization procedures \cite{Giamarchi03} the momentum dependence of 
the interaction is often neglected from the outset (even in the Hamiltonian). In correlation 
functions this leads to the same ultraviolet divergences as described above requiring again 
a regularization `by hand'. Similarly, the purely fermionic approach 
to the single-particle Green function $G^\gtrless(x,t)$ of the TLM by Dzyaloshinski\v{i} 
and Larkin \cite{Dzyaloshinskii73} requires an ad hoc ultraviolet 
regularization. 

We thus emphasize that the integral-free expressions for a variety of 
time-dependent correlation functions which can be found in the literature
cannot be considered as the corresponding {\it exact} correlation functions of the 
TLM. This is often acknowledged by stating that the integral-free expressions of the ad 
hoc procedure only agree to the exact ones at `asymptotically large space-time 
distances'; as discussed in Ref.~\onlinecite{Meden99} (see also below), even this is 
incorrect when considering the decay in the directions specified by 
$x = \pm v t$,
with $v$ being one of the eigenmode velocities at small momentum. 

\subsection{Spectral functions of the Tomonaga-Luttinger model -- the fate of power laws}

\label{sect_fate}
 
We now focus on the two-point correlation function -- the Green function -- 
at temperature $T=0$ from which the single-particle spectral function can be computed by 
Fourier transform.  
The spectral function is of particular importance as it provides forthright access to 
correlation effects and can directly be related to photoemission spectra. We consider the momentum 
integrated function $\rho^{<}(\omega)$ [$\rho^{>}(\omega)$], which is experimentally 
accessible in angular integrated [inverse] photoemission, as well as the momentum resolved 
spectral function $\rho^{<}(k,\omega)$ [$\rho^{>}(k,\omega)$]. A measurement of the latter 
requires momentum resolution. It was shown that the universal low-energy
power-law suppression of $\rho^\gtrless(\omega) \sim |\omega|^\alpha$ for $\omega \to 0$, with
$\alpha>0$ is unaffected by the above ad hoc procedure.\cite{Meden99} Regardless of the 
details of $V(q)$ the exponent $\alpha$ depends on the potential at vanishing momentum 
transfer $V(0)$ (only), that is the constant interaction strength after the ad hoc 
procedure. Furthermore, the power-law nonanalyticity of $\rho^\gtrless(k_{\rm F},\omega)  
\sim |\omega|^{\alpha-1}$ exactly at the Fermi momentum $k=k_{\rm F}$ remains the same; depending 
on the size of $V(0)$ a divergence ($\alpha<1$) or suppression ($\alpha>1$) might 
occur. These findings are consistent with the RG irrelevance of the momentum dependence 
of $V(q)$ as in both cases {\it all} energy scales, that is $\omega$ {\it and,} in the 
case of momentum resolved spectra, $v_{\rm F} (k-k_{\rm F})$ are sent to zero.\cite{Solyom79} 
Here $v_{\rm F}$ denotes the Fermi velocity. 

The question we address here is whether or not any of the standard ad hoc procedures is 
legitimate when it comes to $\rho^\gtrless(k,\omega)$ as a function of $\omega$ at fixed $k-k_{\rm F} \neq 0$. 
Employing these to compute $\rho^\gtrless(k,\omega)$ of the spinless TLM, characteristic 
algebraic threshold nonanalyticities at $\pm v_{c} [k-k_{\rm F}]$ with the charge 
velocity $v_{c}$ of the (collective, bosonic) charge eigenmodes were 
found. In the model with spin 
additional algebraic nonanalyticities appear at $\pm v_{s} [k-k_{\rm F}]$ with the 
velocity $v_{s}$ of the spin modes.\cite{Theumann67,Dzyaloshinskii73,Luther74,Meden92,Voit93} 
The corresponding exponents 
can be expressed in terms of the (momentum independent) interaction potential. For 
small interactions and $k-k_F<0$, $\rho^{<}(k,\omega)$  shows {\it power-law singularities} 
at $\omega =  v_{c} [k-k_{\rm F}]$ and  $\omega =  v_{s} [k-k_{\rm F}]$ instead of a 
single (Lorentzian) peak which would emerge in a Fermi liquid. This is one of the signatures of 
spin-charge separation regarded as a hallmark of (spinful) LLs. In Ref.~\onlinecite{Schoenhammer93} it 
was shown that these features can be found in the exact spectral function of the TLM if a box-like 
potential $V(q) = V(0) \Theta(q_{\rm c}^2 - q^2)$ is assumed, as long as $|k-k_{\rm F}| < q_{\rm c}$ 
with the momentum transfer cutoff $q_{\rm c}$. We emphasize that using a box potential is 
per se not equivalent to the ad hoc procedure, as it can e.g. be seen considering 
$\rho^\gtrless(k,\omega)$ for $|k-k_{\rm F}| >  q_{\rm c}$.\cite{Schoenhammer93} 
For $|k-k_{\rm F}| < q_{\rm c}$ the box potential might, however, 
be viewed as a unique realization of the ad hoc procedure (see below). 
Clearly, a box potential is rather special and might not even be considered as particular 
physical. Thus further work for more generic $V(q)$ is required.   

In Ref.~\onlinecite{Meden99} it was shown that the algebraic properties of the 
Green functions $G^\gtrless(x,t)$ in the space-time plane are significantly affected by the ad hoc procedures. 
It was proven that the exponent of the asymptotic decay of the Green function in the distinguished 
directions $x = \pm v_{c/s} t$ is not only set by $V(0)$ but in addition by a measure 
of the flatness of the potential at $q \to 0$; a result which cannot be obtained within 
any ad hoc procedure. Based on this and the crucial insight that the decay of $G^\gtrless(x,t)$
in the distinguished directions plays a central role in obtaining the power-law 
nonanalyticities in $\rho^\gtrless(k,\omega)$, the question was posed if for generic $V(q)$, $\rho^\gtrless(k,\omega)$ is characterized 
by the `thought to be universal' power laws of the ad hoc procedure. However,  
Ref.~\onlinecite{Meden99} lacks a definite answer. 

The above mentioned box potential is 
`infinitely flat' at $q \to 0$ and thus `nongeneric'. The asymptotics of $G^\gtrless(x,t)$ 
for this and the ad hoc procedures agree and consequently also the features of $\rho^\gtrless(k,\omega)$
at $\omega \approx v_{c/s} [k-k_{\rm F}]$.

It is crucial to realize that a dependence of the line shape of $\rho^\gtrless(k,\omega)$ on the details 
of the interaction away from $q=0$ including such fundamental issues as the presence or absence of algebraic 
nonanalyticities does not contradict the RG irrelevance of the momentum dependence of $V(q)$ in the TLM. 
From this, universality can only be deduced if {\it all} energy scales are sent to zero (see above). 
A fixed $k - k_{\rm F} \neq 0$ sets a scale which becomes active for all generic $V(q)$ 
with $V^{(n)}(q=0) \neq 0$ for some $n \in {\mathbb N}$, where $V^{(n)}(q)$ denotes the $n$-th derivative. 
Thus $\rho^\gtrless(k,\omega)$ cannot expected to be universal on general grounds. In fact, a 
finite scale will destroy the scale invariance of the model and thus `quantum critical' power-law 
scaling. In the case of the at $q=0$ `infinitely  flat' box potential this mechanism is not active as 
long as $|k-k_{\rm F}| < q_{\rm c}$.   

We here supplement Ref.~\onlinecite{Meden99} and provide very strong evidence that the spectral function 
of the TLM at fixed $k - k_{\rm F} \neq 0$ and for a generic potential is not characterized 
by power laws. The latter are only found if $k - k_{\rm F} \to 0$. 
To guide the reader we should from the outset be very precise about our understanding of 
`power-law scaling'. We say that some real function $f$ shows power-law scaling with exponent $\xi$ 
close to $x_0 \in {\mathbb R}$ from above if $d \ln |f(x)| / d \ln(x-x_0)$ approaches $\xi$ for 
$x \to x_0^+$. A similar definition can be given for power-law scaling from below. 
This does of course not exclude that $f$ can to some degree be approximated by a power law or
`resembles' a power law for some range of $x$ (close to $x_0$) even if it does not fulfill 
the above criterion. We will return to this in Sects.~\ref{sect_g4} and 
\ref{sect_g2_g4} in which we present our results for  $\rho^\gtrless(k,\omega)$. 

More generally, we show that even for $|k-k_{\rm F}| \ll q_{\rm c}$ the spectral line 
shape depends on the details of $V(q)$. 

Although the present study might be regarded as somewhat technical -- or 
even pedantic given that issues of the momentum dependence of $V(q)$ in the TLM are virtually 
always nonchalantly ignored -- our results have far reaching consequences.   

\subsection{Implications of our results}

The first implication of our results concerns the concept of LL universality. While $\rho^\gtrless(\omega)$ of 
{\it any} model from the LL universality class shows the power-law suppression of spectral weight for 
$\omega \to 0$ and $\rho^\gtrless(k_{\rm F},\omega)$ a power-law peak or suppression at the same energy, 
{\it LL universality does not predict power laws in}  $\rho^\gtrless(k,\omega)$  for fixed $k - k_{\rm F} \neq 0$.
Evidently, if this type of universality cannot be proven in the low-energy fixed point model, the TLM, 
it cannot be a characteristic feature of the LL universality class.   
This does of course not exclude that {\it certain models} from the LL universality class might show such 
power laws, however, if so for {\it more specific reasons than LL universality.} An obvious example 
for this is the TLM with box potential.\cite{Schoenhammer93} Other examples might be based upon the 
restriction of the (equilibrium) dynamics encountered in specific 1d models with an extensive 
number of local conserved quantities (e.g. the Hubbard model) which are often Bethe ansatz 
solvable.\cite{Essler05} A detailed discussion in 
which we relate our results to spectra of 1d lattice models obtained by either analytical or numerical 
approaches is given in Sect.~\ref{sect_con_LLun}. 

Strongly linked to this are the implications 
of our findings for recent results on $\rho^\gtrless(k,\omega)$ taking the  
nonlinearity of the single-particle dispersion in to 
account\cite{Khodas07,Pereira09,Imambekov09a,Imambekov09b,Schmidt10,Essler15} which are 
embedded in the framework of the so-called 
`nonlinear LL phenomenology'.\cite{Imambekov12} In 
this power laws are not viewed as originating from quantum critical scale invariance but
rather as resulting from a Fermi edge singularity like effect. In the phenomenological 
construction of the effective field theory including curvature effects of the dispersion the above 
described ad hoc regularization is employed. The spectral function is computed for this field theory. 
Our results obtained for {\it linear} LL theory raise the question whether the power laws 
found in `nonlinear LL phenomenology' are robust against a curvature of the bulk two-particle 
potential. We emphasize that these 
power laws are specific to the nonlinearity of the dispersion, which e.g. leads to momentum 
dependent exponents, and are thus different from the nonanalyticities found for the TLM treated 
within the ad hoc procedure (or, for that matter, the TLM with box potential). Again this does 
of course not exclude that for {\it specific 1d models,} e.g. Bethe ansatz solvable lattice 
models, power laws with momentum dependent exponents might be realized. More on this is can be found 
in Sect.~\ref{sect_con_nonlinLL}. 
  
Finally, our findings are of importance for the interpretation of experimental momentum resolved 
spectra. Even after decades of research none of the photoemission experiments on quasi 1d 
metals reporting on the observation of dispersing spin and charge peaks remains unchallenged.\cite{Grioni09} 
One reason for this is that, when interpreting experimental data in the light of LL physics, 
the momentum resolved spectral function obtained within the TLM employing an ad hoc
regularization is taken paradigmatically. Crucially, we find spin and charge peaks
for generic $V(q)$ and a sufficiently small amplitude of the two-particle interaction 
even though they are not given by power-law singularities. Our results show, however, that details 
are model dependent 
(in our case $V(q)$ dependent) and therefore nonuniversal. Thus the detailed spectral features of the 
ad hoc regularized TLM cannot expected to be found in experimental spectra. Further account of the 
relevance of our results for experimental spectra is given in  Sect.~\ref{sect_con_exp}.           
       
\subsection{Structure of the paper}

The rest of our paper is structured as follows. In Sect.~\ref{sect_TLM} we introduce the TLM 
and its bosonization solution. Constructive bosonization of the field operator can be used 
to derive a closed analytical expression for the single-particle Green function, which, however, 
contains a momentum integral on the right hand 
side. This is discussed in Sect.~\ref{sect_Green}. In Sect.~\ref{sect_adhoc} we introduce 
different versions of the ad hoc regularization to obtain integral-free expressions 
for $G^\gtrless(x,t)$. Sections \ref{sect_box} and \ref{sect_arbitrary} are devoted 
to the technical details of how we obtain exact spectra for a box potential and arbitrary 
potentials, respectively. 
In Sect.~\ref{sect_g4} we present results for the spectral function 
of the so-called $g_4$-model with intra-branch interaction only and different shapes 
of momentum dependency of the interaction. These are compared to 
the ones obtained by the ad hoc procedure.  Section~\ref{sect_g2_g4} ist devoted 
to the spectral function of the spinless TLM -- the spinless 
$g_2$-$g_4$-model. In Sect.~\ref{sect_discussion} we 
discuss the implications of our results. When alluding to the relevance 
of our insights for the interpretation of  photoemission data we in addition 
present spectral functions for the spinful TLM.

\section{The Tomonaga-Luttinger model}
\label{sect_TLM}

We here do not introduce the TLM by `deriving' it from the 1d interacting electron gas under certain
assumptions (e.g. on the real-space range of the interaction; no $2k_{\rm F}$ two-particle 
scattering processes)\cite{Schoenhammer05} or as the effective field theory for microscopic 
lattice models\cite{Giamarchi03} but rather take it as a stand-alone model.  
It consists of independent right- ($\alpha=+$) and left-moving ($\alpha=-$) fermions with spin $s$, creation 
operators $a^{\dag}_{k,\alpha,s}$, dispersion $\xi_{\alpha}(k)= 
\alpha v_{\rm F} (k- \alpha k_F)$,
density operators ($q \neq 0$) $\rho_{\alpha,s}(q)= \sum_k 
a_{k,\alpha,s}^{\dagger} a_{k+q,\alpha,s}^{}$, and particle
number operators $n_{k,\alpha,s}=a_{k,\alpha,s}^{\dagger} 
a_{k,\alpha,s}^{}$. Following Luttinger\cite{Luttinger63} an infinite `Dirac sea' filled in the ground state 
is assumed and thus the momentum quantum number $k$ of both particle species is unbounded. This simplifies 
the mathematical treatment as certain relations become operator identities\cite{Haldane81,Schoenhammer05} 
and are not only restricted to the low-energy part of the Hilbert space as in Tomonagas 
approach.\cite{Tomonaga50} This addition of states often requires normal ordering. 
The Hamiltonian for a system of length $L$ is given by
\begin{eqnarray}
&& H  =  \sum_k \sum_{\alpha,s} \xi_{\alpha}(k) 
\left[ n_{k,\alpha,s} - \left< n_{k,\alpha,s}  \right>_0 \right] \nonumber \\
& & + \frac{1}{2 L} \sum_{{q \neq 0} \atop {\alpha, s, s'}} 
\left[g_{4,\parallel}(q) \delta_{s,s'} + 
g_{4,\perp}(q) \delta_{s,-s'} \right]
\rho_{\alpha,s}(q) \rho^{\dag}_{\alpha,s'}(q) \nonumber \\
&&  +   \frac{1}{L} \sum_{{q \neq 0} \atop {s, s'}} 
\left[g_{2,\parallel}(q) \delta_{s,s'} + 
g_{2,\perp}(q) \delta_{s,-s'} \right]
\rho_{+,s}(q) \rho^{\dag}_{-,s'}(q) . 
\label{hamiltonian}
\end{eqnarray}
Here $\left< \ldots \right>_0$ denotes  the 
(noninteracting) ground state expectation value (normal ordering).
We keep the explicit $q$-dependence of the two-particle potential. 
The inter- ($g_2$) and intra-branch ($g_4$) potentials are not 
necessarily equal and replace the potential $V(q)$ referred to in the Introduction. 
Similarly, the interaction of spin parallel ($\parallel$) and anti-parallel 
($\perp$; this is a confusing but standard notation\cite{Solyom79}) particles is not necessarily the same. 
If the TLM is considered as the low-energy fixed point model of the LL universality class 
this flexibility is required. The low-energy physics 
of any model from this class is characterized by four independent numbers, e.g. the 
two LL parameters $K_{c/s}$ and the two velocities 
$v_{c/s}$.\cite{Haldane81,Schoenhammer05} In the TLM for a given $v_{\rm F}$ (and $k_{\rm F}$) 
those are fixed by the (in general) four independent `coupling constants' 
$g_{i,\kappa}(q=0)$ ($i=2,4$; $\kappa=\parallel , \perp$) at 
vanishing momentum transfer (see below). Therefore, to encounter nontrivial interaction effects 
in the LL sense the  $g_{i,\kappa}(q=0)$ should not all be 0. We 
restrict ourselves to these kind of interactions.   
       
We assume that the Fourier transforms $g_{i,\kappa}(q)$ of the two-particle 
interaction are even and for $q \geq 0$  monotonic functions which vanish for 
$|q| \gg q_{\rm c}$, with an interaction cutoff $q_{\rm c}$.  
These requirements are physically sensible if the TLM is considered in its own right.\cite{Tomonaga50}
If, however, the TLM is studied as the effective low energy model it is less clear 
if this assumption holds. It is thus crucial that it the assumption is not essential 
for our main conclusions. Relaxing it would merely complicate the calculations as 
positive and negative momenta would have to be treated seperately.   
We emphasize that at {\it no} stage of the discussion it will be necessary to introduce any 
further ultraviolet cutoffs `by hand' despite the infinite (filled) Dirac sea at negative energies.
In this sense the Hamiltonian Eq.~(\ref{hamiltonian}) represents a mathematically well defined model.

To be more precise the Hamiltonian in Eq.~(\ref{hamiltonian}) defines a {\it whole class of models} 
as the four coupling functions $g_{i,\kappa}(q)$ can be arbitrarily chosen as long as the  
introduced requirements are fulfilled. Still, we continue to refer to this class as {\it the} TLM.
Note that a Hamiltonian of the form Eq.~(\ref{hamiltonian}) but with {\it coupling constants} 
$g_{i,\kappa}$ instead of {\it coupling functions} $g_{i,\kappa}(q)$ can often be found in the literature. 
The necessary ultraviolet regularization is then left implicit and frequently not uniquely defined.  
    
We note that particle number contributions to the Hamiltonian which appear if the model is derived 
from the interacting 1d electron gas\cite{Haldane81,Schoenhammer05} are suppressed as they do not 
play any role for our considerations. 

The spinless
version of the TLM follows from Eq.~(\ref{hamiltonian}) by dropping the spin index and keeping only 
 $g_{i}(q)$ instead of  $g_{i,\kappa}(q)$ for $i=2,4$.

Bosonization of the Hamiltonian and a canonical transformation
lead to\cite{Haldane81,Schoenhammer05}
\begin{eqnarray}
H = \sum_{q \neq 0} \sum_{\nu =c,s} \omega_{\nu}(q) 
\beta_{\nu}^{\dagger}(q) \beta_{\nu}^{}(q) + E_0,  
\label{hamiltoniandiag} 
\end{eqnarray}
with bosonic operators $\beta_{\nu}^{(\dag)}(q)$ describing collective
charge ($\nu=c$) and spin ($\nu=s$) excitations 
(spin-charge separation) as well as the ground state energy $E_0$.
The  $\beta_{\nu}^{(\dag)}(q)$  are linearly related to the densities $\rho_{\alpha,s}(q)$
of the fermions. The mode energies $\omega_{\nu}(q)$ are given by 
\begin{eqnarray} 
\frac{\omega_{\nu}(q)}{|q|}   =  v_{\rm F}  \sqrt{\left( 1+
\frac{g_{4,\nu}(q)}{\pi v_{\rm F}} \right)^{\!\! 2} -
\left( \frac{g_{2,\nu}(q)}{\pi v_{\rm F}}\right)^{\!\! 2}  } 
 = v_{\nu}(q)  ,
\label{energies} 
\end{eqnarray}
where we have introduced the renormalized {\it momentum-dependent} 
charge and spin density velocities $v_{\nu}(q)$ and interactions 
\begin{eqnarray} 
g_{i,c/s}(q) = \left[g_{i,\parallel}(q) 
\pm g_{i,\perp}(q) \right]/2. 
\end{eqnarray}
We already now emphasize that for momentum dependent $g_{i,\kappa}(q)$ the eigenmode 
dispersions will become {\it nonlinear}. The {\it linearization} of the latter,
that is the replacement 
\begin{eqnarray}
 v_{\nu}(q) \to v_{\nu}(0) = v_{\nu},
\label{vcs}
\end{eqnarray}   
is the {\it crucial step} in the ad hoc regularization procedure to derive 
integral-free expressions for correlation functions (see below). 

In Eq.~(\ref{vcs}) we have introduced the $q$-independent charge and spin velocities 
$v_{c/s}$ relevant for the low-energy physics (all energy scales sent to 0) in the LL sense. The
corresponding LL parameters $K_{c/s}$ of the TLM are obtained from 
\begin{eqnarray}
K_{\nu} (q)= \sqrt{\frac{1+g_{4,\nu}(q)/(\pi v_F)-g_{2,\nu}(q)/(\pi
    v_F)}{1+g_{4,\nu}(q)/(\pi v_F)+g_{2,\nu}(q)/(\pi
    v_F)}} 
\label{Kdef}
\end{eqnarray}
in the limit $q \to 0$. Note that $K_{\nu} (q) = 1$ if the inter-branch interaction 
$g_{2,\nu}(q)$ vanishes. 

The spinless version of the bosonized Hamiltonian is obtained after dropping 
the terms with index $\nu=s$.

When presenting our results for $\rho^\gtrless(k,\omega)$ we will initially focus on 
two special cases. The first is 
the spinful TLM with intra-branch interaction only, that is $g_{2,\kappa}(q)=0$, which is 
commonly referred to as the $g_4$-model. The second is the spinless  $g_2$-$g_4$-model.
They are paradigmatic for the two interaction effects characteristic for LLs: spin-charge 
separation and power-law scaling with interaction dependent exponents, respectively.
Proceeding this way increases the transparency of our analysis. 
The scenario for the general spinful model can be deduced by combining the insights of 
both cases; in Sect.~\ref{sect_con_exp} we in addition present a few explicit results 
for the spinful TLM.

\section{The momentum resolved spectral function}

\subsection{The single-particle Green function}
\label{sect_Green}

We are interested in the momentum resolved spectral function. 
It can be computed by Fourier transforms from the greater and lesser 
single-particle Green function
\begin{align}
	i G^>_{\alpha,s} (x,t) & = \langle \psi_{\alpha,s} (x,t) \psi_{\alpha,s}^\dagger (0,0) \rangle \label{Green} \\
	i G^<_{\alpha,s} (x,t) & = \langle  \psi_{\alpha,s}^\dagger (0,0) \psi_{\alpha,s} (x,t) \rangle.
\end{align}
The field operators $\psi_{\alpha,s}^\dagger(x)$ and $\psi_{\alpha,s}(x)$ are related in the usual way 
to the creation and annihilation operators in momentum space
\begin{equation}
	\psi_{\alpha,s} (x)= \frac{1}{\sqrt{L}} \sum_k e^{i k x} a_{k,\alpha,s}.
\end{equation}
Particle-hole symmetry of the TLM ensures\cite{Meden92,Schoenhammer05} 
\begin{equation}
	G^>_{\alpha,s} (x,t) = G^<_{\alpha,s} (-x, -t). 
\end{equation}
and it is sufficient to consider the greater Green function.
We note in passing that band filling is not an issue in the TLM as 
we consider it. However, when the model is investigated as the effective
low-energy model of another 1d correlated electron model the band filling
of the latter will enter in the LL parameters and
velocities characterizing the low-energy physics.\cite{Schoenhammer05} 

Furthermore, with $x \rightarrow -x$ we can go over from the Green function of 
right-movers to the one of left-movers. Therefore, we will only study
$G^>_{+,s}(x,t)$. In the absence of a magnetic field the Green function is spin independent and 
we thus suppress the spin index from now on. 

To compute the ground state expectation value in Eq.~(\ref{Green}) we use constructive bosonization 
of the field operator.\cite{Haldane81,Schoenhammer05} We emphasize that this approach is based on 
operator identities and does not require the introduction of any cutoffs if one first considers 
finite systems of length $L$ (as we will do). In more phenomenological approaches\cite{Giamarchi03} a cutoff 
is introduced -- often denoted by $1/\alpha$ and referred to as an `effective band width' -- which 
formally has to be send to infinity.  However, it is frequently kept finite artificially. This is 
part of one of the possible ad hoc ultraviolet regularizations.         

The exact greater Green function of right-movers for the most general spinful $g_2$-$g_4$-model is given by
\begin{equation}
 i \,G^>_{+} (x,t)  =i  \, \left[ G^>_{+} \right]^0 (x,t) \,  e^{F(x,t)}, 
\label{eq:Ggtrplus_spin}
\end{equation}
with  
\begin{equation}
	  \left[ G^>_{+} \right]^0 (x,t) = \frac{ e^{i (k_{\rm F}+\pi/L) x}}{L} 
\exp \left\{ \sum_{n=1}^\infty \frac{1}{n}  e^{ i q_n (x- v_{\rm F} t + i 0^+)} \right\}.
\label{freeGreen}
\end{equation}
The interaction enters in 
\begin{align}
F(x,t)  = & \frac{1}{2} \sum_{\nu = c,s} \sum_{n=1}^\infty \frac{1}{n}  
\left[ e^{ i  q_n x } \left( e^{- i \omega_\nu(q_n) t} - e^{ - i v_{\rm F} q_n t} \right) \right. 
\nonumber \\
& + \left. 2 \gamma_\nu(q_n) 
\left( \cos (q_n x) e^{- i \omega_\nu(q_n)  t} -1  \right) \right], 
\label{eq:F_g2g4spinful}
\end{align}
with
\begin{eqnarray}
\gamma_{\nu}(q)=\left[ K_{\nu}(q)+ 1/K_{\nu}(q) -2 \right]/4 
\label{gammadef} 
\end{eqnarray}
and $q_n=n 2\pi/L $ (periodic boundary conditions).
Due to the decay of the $g_{i,\kappa}(q)$ on the scale $q_{\rm c}$ the  momentum sum in 
$F(x,t)$  is convergent in the ultraviolet. 
Note in particular that for $q_n \gg q_{\rm c}$ the two terms in the first line cancel each other as 
$\omega_\nu(q_n) \to  v_{\rm F} q_n$ in this limit [compare Eq.~(\ref{energies})]. 
The term in the second line of Eq.~(\ref{eq:F_g2g4spinful}) is convergent as 
$\gamma _\nu(q_n) \to 0$ for   $q_n \gg q_{\rm c}$ [compare Eqs.~(\ref{gammadef}) and (\ref{Kdef})]. 
For vanishing interaction $F(x,t)=0$. Thus $\left[ G^>_{+} \right]^0 (x,t)$ is the noninteracting Green function.   
In the thermodynamic limit $L \to \infty$ it becomes 
\begin{equation}
	  \left[ G^>_{+} \right]^0 (x,t) =  \frac{1}{2 \pi} \frac{e^{i k_{\rm F} x}}{x-v_{\rm F}t +i 0^+} .
\label{exactGreen}
\end{equation} 
The factor $\exp{(- q_n 0^+)}$ in Eq.~(\ref{freeGreen}) ensuring convergence appears naturally, and is 
not related to any ad hoc regularization. Only with this factor the exactly known $ \left[ G^>_{+} \right]^0 (x,t)$ 
Eq.~(\ref{exactGreen}) and from this the exact noninteracting spectral function 
\begin{eqnarray}
 \left[ \rho^>_{+} (k, \omega ) \right]^0 (x,t) = \Theta(k-k_{\rm F}) \delta(\omega-\xi_{+}(k))  
\end{eqnarray}
can be obtained. Due to the linear single-particle dispersion $ \left[ G^>_{+} \right]^0 (x,t)$  
is of relativistic form.  
The greater spectral function is defined as
\begin{equation}
\rho^>_{+} (k_n, \omega ) = \frac{1}{2 \pi} \int_{-\infty}^\infty 
dt \, e^{i \omega t} \int_{-L/2}^{L/2} dx \, e^{- i k_n x} \; i G^>_{+} (x,t) .
\end{equation}
As $G^>_{+} (x,t)$ Eqs.~(\ref{eq:Ggtrplus_spin}) to (\ref{eq:F_g2g4spinful}) is an analytic function 
in the lower half of the complex $t$-plane $\rho^>_{+} (k, \omega ) $
has nonvanishing weight only for $\omega \geq 0$. 

To compute $\rho^>_{+} (k, \omega )$ for arbitrary potentials in the thermodynamic limit 
three nested integrals have to be performed. The integrands are slowly decaying, 
oscillatory, and have poles close to the real axis. Therefore, the accuracy which can 
be achieved when straightforwardly performing the integrals numerically, given $g_{i,\kappa}(q)$, is 
not sufficient to answer the question whether or not power laws can be found for 
$k-k_{\rm F} \neq 0$. We are thus forced to proceed differently. Before presenting 
our approach to the exact spectral function of the TLM in Sects.~\ref{sect_box} and 
\ref{sect_arbitrary} we will describe approximate ad hoc procedures which were pursued in
the literature to make analytical progress. 
   
\subsection{The ad hoc regularization}
\label{sect_adhoc}

In the natural way  of writing the {\it exact} Green function of 
the TLM Eq.~(\ref{eq:Ggtrplus_spin}) the noninteracting one is factorized out.\cite{Meden99} However, 
other expressions 
for $G^>_{+} (x,t)$ can be found in the literature.\cite{Luther74,Meden92,Voit93,Voit95} These 
result from the following procedure. The term in the curly brackets of Eq.~(\ref{freeGreen}) 
is canceled against the last term in the first line of Eq.~(\ref{eq:F_g2g4spinful}). 
After this step the remaining $q$-sum in Eq.~(\ref{eq:F_g2g4spinful}) is no longer 
convergent in the ultraviolet. Two ways have been reported how to deal with this problem. 

In the first (i) the remaining first term of the first line of  Eq.~(\ref{eq:F_g2g4spinful}) is 
multiplied by $\exp{(- q_n 0^+)}$. To obtain integral-free expressions for $G^>_{+} (x,t)$ one then 
assumes that $\gamma_{\nu}(q_n)$ is given by $\gamma_{\nu} \exp{(- q_n \Lambda)}$ which can be reached 
by choosing a proper momentum dependence of the $g_{i,\kappa}(q)$. Thus one simply 
selects a certain interaction potential. In {\it addition,} one linearizes the 
eigenmode dispersions $\omega_{\nu}(q) \to v_\nu q$ for all $q$. This is {\it not}
a consequence of the special choice of the interaction potential but is rather an approximation 
done independently `by hand'. 
In the thermodynamic limit one then obtains
\begin{align}
 [G_{+}^{>}]_{\rm (i)} (x,t)   = &    \frac{e^{i k_{\rm F} x}}{2 \pi}\prod_{\nu =c,s} 
\left[ \frac{1}{x- v_{\nu} t +i 0^+ } \right]^{1/2} \nonumber \\* 
& \times \left[ \frac{\Lambda^2}
{\left( x- v_{\nu}  t + i \Lambda \right)
 \left( x+ v_{\nu}  t - i \Lambda \right) } 
\right]^{\gamma_{\nu}/2} .
\label{gapprox1}
\end{align}   

In the second ad hoc procedure (ii) the entire (remaining) 
argument of the sum in Eq.~(\ref{eq:F_g2g4spinful}) is multiplied by $\exp{(- q_n \Lambda)}$ with 
a {\it finite} momentum cutoff $1/\Lambda>0$. The momentum dependence of $\gamma_{\nu}(q_n)$ is dropped 
and $\omega_{\nu}(q)$ is again linearized. 
This leads to 
\begin{align}
 [G_{+}^{>}]_{\rm (ii)} (x,t)   = &    \frac{e^{i k_{\rm F} x}}{2 \pi}\prod_{\nu =c,s} 
\left[ \frac{1}{x- v_{\nu} t +i \Lambda } \right]^{1/2} \nonumber \\* 
& \times \left[ \frac{\Lambda^2}
{\left( x- v_{\nu}  t + i \Lambda \right)
 \left( x+ v_{\nu}  t - i \Lambda \right) } 
\right]^{\gamma_{\nu}/2} .
\label{gapprox2}
\end{align}

Finally, a third way (iii) to obtain an integral-free expression for $G^>_{+} (x,t)$ is based directly 
on  Eqs.~(\ref{eq:Ggtrplus_spin}) to (\ref{eq:F_g2g4spinful}). In this the 
momentum dependence of $\gamma_{\nu}(q_n)$ is dropped and $\omega_{\nu}(q)$ is linearized after the 
square bracket in Eq.~(\ref{eq:F_g2g4spinful}) was multiplied by $ \exp{(- q_n \Lambda)}$.
This way one obtains
\begin{align}
[G_{+}^{>}]_{\rm (iii)} (x,t)   =  &  \frac{e^{i k_{\rm F} x}}{2 \pi} \frac{1}{x- v_{\rm F}  t +i0^+} \nonumber \\* 
& \times  \prod_{\nu =c,s} 
\left[ \frac{x- v_{\rm F}  t + i \Lambda}
{x- v_{\nu} t + i \Lambda } \right]^{1/2}  \nonumber \\* 
& \times 
\left[ \frac{\Lambda^2}
{\left( x- v_{\nu}  t + i \Lambda \right)
 \left( x+ v_{\nu}  t - i \Lambda \right) } 
\right]^{\gamma_{\nu}/2} .
 \label{gapprox3}
\end{align}
We note that for the special case $\gamma_s=0$ (spin-rotational invariant interaction) and 
$v_s = v_{\rm F}$ this is exactly the expression derived by  Dzyaloshinski\v{i} and 
Larkin \cite{Dzyaloshinskii73} in a purely fermionic approach which is  based on Ward identities 
and the closed loop theorem. \cite{Bohr81}   

Obviously, all three {\it approximate} functions differ. However, their asymptotic behavior 
(power-law decay) for large space-time arguments is the same, in particular along the four 
special directions $x = \pm v_\nu t$. For this reason they all lead to the same  
power laws in $\rho^>_{+} (k, \omega )$ for $\omega$ close to $\pm v_{\nu} (k-k_{\rm F})$ 
as derived in Refs.~\onlinecite{Meden92,Voit93,Voit95,Meden99}. 

Similarly, for $x=0$ all three approximate expressions
have the same asymptotic behavior at large $|t|$, $\sim t^{-\alpha-1}$, 
with $\alpha = \gamma_{c} + \gamma_{s}$ (and are analytic functions in the 
lower half of the complex $t$-plane). After a single Fourier transform from time 
to frequency this gives $\rho^{>}(\omega) \sim \Theta(\omega) \omega^\alpha$. 
As discussed in Ref.~\onlinecite{Meden99} this power-law scaling of the momentum integrated spectral 
function is also found based on the {\it exact} expressions (\ref{eq:Ggtrplus_spin}) to 
(\ref{eq:F_g2g4spinful}) and for arbitrary $g_{i,\kappa}(q)$ (fulfilling the above mentioned restrictions) 
with  $\gamma_{\nu} \to \gamma_{\nu}(q=0)$. This universality is based on  
the RG irrelevance of the momentum dependence of the interaction;\cite{Solyom79} 
the appearance of the power law with an exponent set by the two-particle potential 
at vanishing momentum transfer $q=0$ is not affected by the potential away from this point. 
Any of the discussed ad hoc procedures (i) to (iii) can thus be employed without spoiling 
the universal behavior. The same holds at $t=0$ but $x \neq 0$ which after Fourier transform
leads to the momentum distribution function $n_+(k)$ which also shows universal power-law scaling 
for $k \to k_{\rm F}$.\cite{Voit95,Giamarchi03,Schoenhammer05}     
The question we address here is whether or not the ad hoc regularized Green functions 
can also be used to obtain universal results for $\rho^>_{+} (k, \omega )$ at $k-k_{\rm F} \neq 0$. 

As already emphasized the linearization of the $\omega_{\nu}(q)$ is the crucial step which leads 
to power laws in $\rho^>_{+} (k, \omega )$ for $k - k_{\rm F} \neq 0$ in the ad hoc procedures. 
In Ref.~\onlinecite{Meden99} it was shown using a generalization of the stationary phase method that 
any $q=0$-curvature of $v_\nu(q)$ affects the asymptotic behavior of  $G^>_{+} (x,t)$ in the 
distinguished directions $x = \pm v_\nu t$. 
The decay is no longer solely given by the $\gamma_\nu$ and therefore not only by the $g_{i,\kappa}(0)$. 
However, in the ad hoc procedures (with constant $v_\nu$), one can analytically
show that it is the asymptotic behavior in these directions of the $x$-$t$-plane which leads to the 
power laws at $\omega = \pm v_\nu [k-k_{\rm F}]$.\cite{Meden99} This raises doubts that the latter 
are generic.

Besides the spinful TLM we also consider its spinless version. From the above expressions 
for $ G_{+}^{>} (x,t)$ [including the ones of the ad hoc regularization (i) to (iii)] the 
spinless Green function is obtained by setting $\gamma_c=\gamma_s$, 
$\omega_c=\omega_s$, and $v_c=v_s$. All what was said in the last two paragraphs 
about power-law behavior and universality  remains valid in the spinless case up to 
the (obvious) exception that in the ad hoc regularized spectral function only the 
two (instead of four) distinguished energies  $\pm v_{c} (k-k_{\rm F})$ exist.  

We note in passing that it was realized decades ago that within the
ad hoc procedures (i) and (ii) exact spectral sum rules are not 
fulfilled.\cite{Suzumura80,Schoenhammer93b,Voit93b}

\subsection{How to compute $\rho^>_{+} (k, \omega )$ for a box potential}
\label{sect_box}

In the Introduction we mentioned that exact results for the momentum resolved spectral function 
of the TLM were derived based on Eqs.~(\ref{eq:Ggtrplus_spin}) to 
(\ref{eq:F_g2g4spinful}) assuming a box potential.\cite{Schoenhammer93} We will compare our results 
for other potentials to these. To be self-contained we here give all the formulas required to obtain 
$\rho^>_{+} (k, \omega )$ of the spinful $g_4$-model and the spinless $g_2$-$g_4$-model 
for $g_{i,\kappa}(q)= g_{i,\kappa} \Theta(q_{\rm c}^2 - q^2)$. More details are presented for the first 
case which was not separately discussed in Ref.~\onlinecite{Schoenhammer93}. We note 
that for a reader primarily interested in results it is not necessary to understand the 
technical details of this section in full detail.  

\subsubsection{The spinful $g_4$-model}

\label{sping4box}

For $g_{2,\kappa}(q)=0$ it directly follows that $\gamma_\nu(q)=0$ as $K_\nu(q)=1$; 
compare Eqs.~(\ref{gammadef}) and (\ref{Kdef}). Equations  
(\ref{eq:Ggtrplus_spin}) to (\ref{eq:F_g2g4spinful}) for the exact Green function
thus simplify considerably. The same holds for the ad hoc regularized 
expressions Eqs.~(\ref{gapprox1}) to (\ref{gapprox3}) as $\gamma_\nu =0$. 
These are characterized by square-root singularities at $x=v_c t$ and 
$x=v_s t$. After Fourier transform they lead to square root singularities in 
$\rho^>_{+} (k, \omega )$ for 
$\omega \to v_c [k-k_{\rm F}]$ and $\omega \to v_s [k-k_{\rm F}]$.\cite{Meden92,Voit93,Voit95}  
We note that $[G_{+}^{>}]_{\rm (iii)} (x,t)$ 
contains the additional factor  $(x- v_{\rm F}  t + i \Lambda)/(x- v_{\rm F}  t +i0^+)$ 
which does not drop out as $\Lambda>0$. Further down we will discuss how this term affects 
the spectral properties.  As the ground state of the $g_4$-model remains 
the noninteracting one,\cite{Schoenhammer93} 
$\rho^>_{+} (k, \omega )$  has finite weight only for $k \geq k_{\rm F}$. 

For the case of a box potential, it is 
\begin{equation}
	v_\nu(q) = 
	\begin{cases}
		v_\nu, & q \leq q_{\rm c} \\
		v_{\rm F}, & q > q_{\rm c} .
	\end{cases}
\end{equation}
Setting for convenience 
$z = \exp \{ i (2 \pi / L ) ( x - v_{\rm F} t) \}$ as well as 
$z_\nu = \exp \{ i (2 \pi / L ) ( x - v_\nu t) \}$ and expanding the exponential function one obtains
from Eqs.~(\ref{eq:Ggtrplus_spin}) to (\ref{eq:F_g2g4spinful})
\begin{align}
	i G^>_{+} (x,t)  & =  \frac{1}{L} \,  
	e^{i (2 \pi/L) (n_{\rm F}+1) x } \:  \left( \sum_{l=0}^\infty z^l \right)   \notag \\
	&  \times \prod_{n=1}^{n_{\rm c}} \left( \sum_{m=0}^\infty \frac{ (-1/n)^m }{m!} z^{n m} \right) 
	\left( \sum_{j=0}^\infty \frac{ (1/(2n))^j }{j!} z_c^{n j} \right)  \notag \\
	&  \times \left( \sum_{l=0}^\infty \frac{ (1/(2n))^l }{l!} 
	z_s^{n l} \right) \label{eq:g4spinG} \\
	& \overset{!}{=} \frac{1}{L} \,  e^{i (2 \pi/L) (n_{\rm F}+1) x } \: \left( \sum_{m=0}^\infty a_m^{(n_{\rm c})} z^m \right) 
 \notag \\
	&  \times 
	\left( \sum_{j=0}^\infty b_j^{(n_{\rm c})} z_c^j \right) \left( \sum_{l=0}^\infty b_l^{(n_{\rm c})} 
	z_s^l \right), \label{eq:g4spinG_coeff}
\end{align}
with $q_{\rm c}= n_{\rm c} 2\pi/L$. 
Here $k_{\rm F} = (2 n_{\rm F} + 1) \pi /L$ and $n_{\rm F}$ is the index of the last occupied 
single-particle state; $k_{\rm F}$ thus lies in between the last occupied and the first unoccupied 
one.
The coefficients in Eq.~\eqref{eq:g4spinG_coeff} can be determined by a recursion relation, 
where $m > 1$, $l \in \mathbb{N}_0$ and $i = 0, \dots, m-1$:
\begin{align}
	& a_{lm+i}^{(m)} = \sum_{j=0}^l \frac{ (-1/m)^j}{j!} \ a_{m(l-j)+i}^{(m-1)} , \label{eq:a_rec} \\
	& b_{lm+i}^{(m)} = \sum_{j=0}^l \frac{ (1/(2m))^j}{j!} \ b_{m(l-j)+i}^{(m-1)} .
\end{align}
The initial values are given by 
\begin{equation}
	a_m^{(1)}  = \sum_{j=0}^m \frac{ (-1)^j }{ j! } \label{eq:a_rec_ini}
\end{equation}
and $b_m^{(1)} = (1/2)^m/m!$. The recursion can easily be performed on a computer. 
The double Fourier transform can be computed analytically and one obtains
\begin{align}
	\rho^>_{+} (k_n, \omega) & = \sum_{l=0}^{\tilde n} \sum_{j=0}^{\tilde n-l} 
	a_{\tilde n-l-j}^{(n_{\rm c})} \, b_l^{(n_{\rm c})} 	\, b_j^{(n_{\rm c})}  \notag \\
	& \times 
	\delta \left[ \omega - \frac{2 \pi }{L} \big( (\tilde n-l-j) v_{\rm F} + l v_c + j v_s \big) \right] ,
\label{sping4rec}
\end{align}
with $\tilde n= n-(n_{\rm F} +1)$. This way the exact spectral function $\rho^>_{+} (k_n, \omega)$ of the TLM with
box potential can easily be computed for large but finite systems (see Sect.~\ref{sect_g4}). 

To obtain analytical insights we rewrite the Green function as
\begin{align}
	i G^>_{+} (x,t)  &  =  \frac{1}{L} \,  e^{i (2 \pi/L) (n_{\rm F}+1) x }   \notag \\
& \times 
\exp \left\{ \frac{1}{2} \sum_{n=1}^{n_{\rm c}} \left( \frac{ z_c^n}{n} + \frac{ z_s^n }{n} \right) + 
\sum_{n=n_{\rm c}+1}^\infty \frac{z^n}{n} \right\}.
\end{align}
Thus
\begin{equation}
	\sum_{m=0}^\infty a_m^{(n_{\rm c})} z^m = 1 + \sum_{n=n_{\rm c}+1}^\infty \frac{z^n}{n}  
+ \frac{1}{2} \left( \sum_{n=n_{\rm c}+1}^\infty \frac{z^n}{n}  \right)^2 + \ldots
\end{equation}
and we immediately see that $a_0^{(n_{\rm c})} = 1$, $a_m^{(n_{\rm c})} = 0$ for $1 \leq m \leq n_{\rm c}$ and $a_m^{(n_{\rm c})} 
= 1/m$ for $ n_{\rm c}+1 \leq  m \leq 2 n_{\rm c} +1$.
For $\tilde{n} \leq n_{\rm c}$ the simplified expression
\begin{equation}
	\rho^>_{+} (k_n, \omega) 
= \sum_{l=0}^{\tilde{n}} b_l^{(n_{\rm c})} b_{\tilde{n}-l}^{(n_{\rm c})} \ \delta \left[ \omega -
\frac{2 \pi }{L} \left( \tilde{n} v_s + l (v_c - v_s) \right) \right] 
\label{eq:g4spinrhoex}
\end{equation}
holds. For fixed $\tilde{n}$ and $v_s < v_c$, there is only spectral weight for 
$v_s [k-k_{\rm F}]  \leq \omega \leq v_c [k-k_{\rm F}]$ (up to corrections of order $1/L$). 
For $v_s > v_c$ the roles of the two velocities are interchanged. 
We note that for $k-k_{\rm F} \leq q_{\rm c}$ the bare Fermi velocity $v_{\rm F}$ 
drops out.   

Further analytical results can be obtained employing 
\begin{align}
&	\exp \left\{ \alpha \sum_{n=1}^{n_{\rm c}} \frac{1}{n} x^n \right\}  = (1-x)^{-\alpha} \, \exp \left\{- \alpha \sum_{n=n_{\rm c}+1}^\infty \frac{1}{n} x^n \right\} \notag \\
	& = \left[ \sum_{j=0}^\infty \binom{-\alpha}{j}  (-x)^j \right] \, \exp \left\{- \alpha \sum_{n=n_{\rm c}+1}^\infty \frac{1}{n} x^n \right\}, \label{eq:Identity_expgammasum}
\end{align}
where $\binom{-\alpha}{j}$ is the generalized binomial coefficient. 
From this it follows that $b_0^{(n_{\rm c})} = 1$ and 
\begin{equation}
	b_j^{(n_{\rm c})} = (-1)^j \binom{-1/2}{j} \quad \xrightarrow{1 \ll j \leq n_{\rm c}} \: \text{const.} \times j^{-1/2},
\end{equation}
where
\begin{equation}
	(-1)^j \binom{-\alpha}{j} 
\approx \frac{1}{ \Gamma[\alpha] } j^{\alpha-1} \quad \text{for }  j \rightarrow \infty
\end{equation}
was used.
Inserting this for energies close to $ v_s [k_n-k_{\rm F}]$ into Eq.~\eqref{eq:g4spinrhoex}, where 
$l$ is the integer next to $(\omega - v_s [k_n -k_{\rm F}] ) / [(2 \pi/L) (v_c - v_s)]$, we obtain for 
$L \rightarrow \infty$ the one-sided square-root singularity ($v_c > v_s$)
\begin{equation}
	\rho^>_{+} (k, \omega) \sim \Theta 
\left( \omega- v_s [k-k_{\rm F}] \right) 
\left( \omega - v_s [k-k_{\rm F}] \right)^{-1/2}.
\label{wurz}
\end{equation}
Analogously we get close to $ v_c [k_n-k_{\rm F}]$
\begin{equation}
	\rho^>_{+} (k, \omega) \sim \Theta 
\left( \omega- v_c [k-k_{\rm F}] \right) 
\left(\omega - v_c [k-k_{\rm F}] \right)^{-1/2}.
\end{equation}
For $v_s > v_c$ we only need to interchange the two velocities.

We have thus shown that the exact spectral 
function of the $g_4$-model with box potential shows the edge singularities also found within 
the ad hoc procedures (i) to (iii). 
 
We note in passing that for $k=k_{\rm F}$ the spectral function reduces to a $\delta$-function of
weight 1 located at $\omega=0$.   

\subsubsection{The spinless $g_2$-$g_4$-model}

\label{nospinbox}

In Sect. IV of Ref.~\onlinecite{Schoenhammer93} a recursive way of computing the spectral function for the 
full spinless TLM with box potential similar to Eqs.~(\ref{eq:a_rec})-(\ref{sping4rec})  was introduced. 
It is given by 
\begin{align}
\rho^>_+(k_n, \omega) & = A^{-2 \gamma_c}
\sum_{r=\max\{0, -\tilde{n}\}}^\infty \sum_{l=0}^{\tilde{n}+r}  a_{\tilde{n}+r-l}^{(n_{\rm c})} b_l^{(n_{\rm c})} 
c_r^{(n_{\rm c})} \notag \\ 
& \times  \delta \left[ \omega - \frac{2 \pi}{L} \big( ( \tilde{n}+r-l ) v_{\rm F} + 
(r+l) v_c \big) \right],
	\label{eq:g2g4rho_boxex}
\end{align}
with  $A = \exp\{ \sum_{n=1}^{n_{\rm c}} (1/n) \}$.  
The coefficients $a_m^{(n_{\rm c})}$ are determined as in Eq.~\eqref{eq:a_rec} and \eqref{eq:a_rec_ini}. 
For $b_m^{(n_{\rm c})}$, the recursion relation reads
\begin{equation}
	b_{lm+i}^{(m)} = \sum_{j=0}^l \frac{1}{j!} \left( \frac{1+\gamma_c}{m} \right)^j \ b_{m(l-j)+i}^{(m-1)} ,
\end{equation}
with the initial values $b_m^{(1)} = (1+\gamma_c)^m/m!$, and for the $c_m^{(n_{\rm c})}$ 
\begin{equation}
	c_{lm+i}^{(m)} = \sum_{j=0}^l \frac{1}{j!} \left( \frac{\gamma_c}{m} \right)^j \ c_{m(l-j)+i}^{(m-1)} ,
\end{equation}
with the initial values $c_m^{(1)} = \gamma_c^m/m!$. 

In analogy to the considerations for the spinful $g_4$-model, we can infer from 
the behavior of the coefficients the behavior of the spectral function close to 
$v_c [k-k_{\rm F}]$ (for $L \to \infty$).
For $k-k_{\rm F} >0 $ we find\cite{Schoenhammer93}  
\begin{equation}
	\rho^>_+(k, \omega) \sim \Theta( \omega - v_c [k-k_{\rm F}]  ) \ 
\left( \omega - v_ c[k-k_{\rm F}]  \right)^{\gamma_c -1},
\label{power1}
\end{equation}
that is a divergence if $\gamma_c <1$ (not to strong interactions).  
In contrast to the $g_4$-model $\rho^>_+$ can now also carry spectral weight 
for $(k-k_{\rm F}) <0 $. At the threshold we obtain\cite{Schoenhammer93} 
\begin{equation}
	\rho^>_+(k, \omega) \sim \Theta( -\omega - v_c [k-k_{\rm F}] ) \ 
\left( -\omega - v_c [k-k_{\rm F}]  \right)^{\gamma_c},
\label{power2}
\end{equation}
that is a power-law suppression since $\gamma_c  > 0$ in the full model. 
In the special case $k=k_{\rm F}$, power-law behavior with $\rho^>_+(k_{\rm F}, \omega) 
\propto \omega^{2 \gamma_c -1}$ is found. These threshold power laws can also be found 
based on the Green function of the ad hoc procedures (i) to (iii) discussed in 
Sect.~\ref{sect_adhoc}.\cite{Theumann67,Luther74,Meden92,Voit93}

\subsection{How to compute $\rho^>_{+} (k, \omega )$ for arbitrary potentials}
\label{sect_arbitrary}

We now show that expressions for the Green and the spectral function which involve 
recursively computed coefficients can also be given for an arbitrary 
momentum dependence of the two-particle potential.
For the box potential the dispersion of the elementary charge and spin modes 
is piecewise linear. This changes for arbitrary potentials. As a consequence 
the coefficients become time-dependent and the Fourier transform with respect to time
has to be performed numerically, e.g. using fast Fourier transform (FFT).    

\subsubsection{The spinful $g_4$-model}

\label{sping4arb}

For arbitrary potentials the spin and charge velocity are no longer piecewise 
momentum independent. If the potential is effectively zero for $ q > \tilde{q}_{\rm c}$, 
with a $\tilde{q}_{\rm c}$ which we take sufficiently larger then the characteristic scale 
$q_{\rm c}$, we can work with $g_{4,\nu} (q) \rightarrow g_{4,\nu} (q) \, \Theta(\tilde q_{\rm c}^2 - q^2)$
for all practical purposes. Then we can rewrite $F(x,t)$ in Eq.~(\ref{eq:F_g2g4spinful})   
\begin{align}
F(x,t) & = \sum_{n=1}^{\tilde{n}_{\rm c}} \frac{1}{n} \Bigg( \frac{1}{2} 
e^{-i \omega_c(q_n) t} +\frac{1}{2} e^{-i \omega_s(q_n) t}   \notag \\ 
& - e^{-i (2 \pi/L) n v_{\rm F} t} \Bigg) \, z^n ,
\end{align}
with $ z = \exp \{ i (2 \pi/L) x \}$, and use \textit{time-dependent} coefficients 
to write 
\begin{align}
i G^>_{+} (x,t) &= \frac{1}{L} \,  e^{i (2 \pi/L) (n_{\rm F}+1) x } \:  
\left( \sum_{l=0}^\infty z^l \, e^{i (2 \pi/L) l v_{\rm F} t }  \right)   \notag \\
	&  \times \prod_{n=1}^{\tilde{n}_{\rm c}} \Bigg( \sum_{m=0}^\infty \frac{(1/n)^m}{m!} 
\Bigg[ \frac{1}{2}  e^{- i \omega_c(q_n) t} \notag \\ &  +\frac{1}{2} e^{-i \omega_s(q_n) t}-  
e^{-i  (2 \pi/L) n v_{\rm F} t} \Bigg]^m z^{m n}\Bigg)  \\
	& \overset{!}{=} \frac{1}{L} \,  e^{ i (2 \pi/L) (n_{\rm F}+1) x } \sum_{m=0}^\infty a_m^{(\tilde{n}_{\rm c})} (t) 
\, z^m.
\end{align}
The recursion relation for the $a_m^{(\tilde{n}_{\rm c})}(t)$ is given by
\begin{align}
& a_m^{(1)} (t)  = \sum_{l=0}^m \frac{1}{l!} \left[ e^{-i(2 \pi/L) v_{\rm F} t }  \right]^{m-l} \notag \\
& \times 
\left[ \frac{1}{2}  e^{-i  \omega_c(q_1)    t } + \frac{1}{2}  e^{-i  \omega_s(q_1)  t } -  
 e^{-i(2 \pi/L) v_{\rm F}  t } \right]^l , \\
& a_{lm+i}^{(m)} (t)  = \sum_{j=0}^l \frac{(1/m)^j}{j!}  \Bigg[ \frac{1}{2} e^{-i  \omega_c(q_m) t } \notag \\ & +
 \frac{1}{2}  e^{-i  \omega_s(q_m)  t } -  e^{-i(2 \pi/L) m v_{\rm F}  t } \Bigg]^j \ a_{m(l-j)+i}^{(m-1)}(t),
\end{align}
where in the second line $m > 1$, $l \in \mathbb{N}_0$ and $i = 0, \dots, m-1$. 
With this representation of the Green function, the Fourier transform to momentum space 
can be performed analytically
\begin{equation}
i G^>_{+} (k_n,t) = \frac{1}{2 \pi} a_{\tilde{n}}^{(\tilde{n}_{\rm c})} (t). 
\label{eq:g4spinG(k,t)}
\end{equation}
The remaining Fourier transform 
 \begin{equation}
\rho^>_{+} (k_n, \omega) = \int_{-\infty}^\infty d t \ e^{i \omega t} \ i G^>_+ (k_n, t)  
\end{equation}
can be performed numerically as a FFT. Since for the finite system the spectral function 
consists of $\delta$-peaks, the Green function does not decay in time. Therefore, we have to multiply 
$i G^>_+(k_n,t)$ with a decaying function before performing the FFT. Here, we will always use the 
exponential function $\exp \{ -\chi |t| \}$. In frequency space, this corresponds to a convolution of 
the spectral function with the Lorentzian  $\pi^{-1} \chi / ( \omega^2 + \chi^2) $, i.e. each $\delta$-peak 
is broadened into a Lorentzian of width $\chi$.

\subsubsection{The spinless $g_2$-$g_4$-model}
\label{nospinarb}

In the same way as for the spinful $g_4$-model, we can introduce recursively defined time-dependent 
coefficients for the spinless $g_2$-$g_4$-model
\begin{align}
a_m^{(1)} (t) & = \sum_{l=0}^m \frac{1}{l!} \left[ e^{-i (2 \pi/L) v_{\rm F} t} \right]^{m-l} \notag \\
& \times \left[ \left( 1+\gamma_c(q_1) 
\right) e^{-i  \omega_c(q_1)  t} -  e^{-i (2 \pi/L) v_{\rm F} t} \right]^l \\
a_{lm+i}^{(m)} (t) & = \sum_{j=0}^l \frac{ (1/m)^j }{j!} \Big[ \left( 1+\gamma_c(q_m) \right) 
e^{-i  \omega_c(q_m)  t} \notag \\ & -  e^{-i (2 \pi/L) m v_{\rm F} t} \Big]^j \ a_{m(l-j)+i}^{(m-1)} (t) 
\end{align}
and
\begin{align}
& b_m^{(1)} (t)  = \frac{1}{m!}  \left[  \gamma_c(q_1) e^{-i  \omega_c(q_1) t}  \right]^m \\
& b_{lm+i}^{(m)} (t) = \sum_{j=0}^l \frac{ (1/m)^j }{j!} \left[ \gamma_c(q_m) e^{-i   \omega_c(q_m)  t}  \right]^j 
\ b_{m(l-j)+i}^{(m-1)} (t).
\end{align}
In terms of these we can rewrite
\begin{align}
i G^>_+(k_n, t) & = \frac{1}{2 \pi} \exp \left\{ -2 \sum_{n=1}^{\tilde{n}_{\rm c}} \frac{\gamma_c(q_n)}{n} \right\} 
\notag \\ & \times
\sum_{r=\max \{0, -\tilde{n} \}}^{\infty} a_{ \tilde{n}+r }^{ (\tilde{n}_{\rm c}) } (t)  \, 
b_r^{ (\tilde{n}_{\rm c}) } (t) 
\label{eq:g2g4spinless_G_coeff}
\end{align}
and the remaining Fourier transform to obtain $\rho^>_{+} (k_n, \omega)$  can be performed, again 
after multiplication with $\exp\{-\chi |t| \}$, numerically by means of a FFT.   

\subsection{Spectra of the spinful $g_4$-model}
\label{sect_g4}

Based on the formulas given in Sects.~\ref{sping4box} and \ref{sping4arb} we are in a position to compute 
the exact spectral function of the spinful $g_4$-model for different potentials at finite system size $L$. 
For the box potential both Fourier transforms can be performed analytically while for arbitrary 
potentials the time-transform is performed numerically as a FFT. In this case only broadened spectra
can be obtained. We will compare the results to those derived from one of the ad hoc procedures  
of Sect.~\ref{sect_adhoc}. 

Besides the box potential we consider ($\nu=c,s$)
\begin{align}
	g^{\rm p=4}_{4,\nu}(q) &  =  g_{4,\nu} \exp\{ -(q/q_{\rm c})^4/9\} \ & \text{ (p=4),} \label{p4pot} \\
	g^{\rm gauss}_{4,\nu}(q) &  = g_{4,\nu} \exp\{ -(q/q_{\rm c})^2 \} \ & \text{ (gauss),} 
        \label{eq:pot_gaus} \\
	g^{\rm exp}_{4,\nu}(q) & = g_{4,\nu} \exp\{ -3 \, |q/q_{\rm c}| \} \ & \text{ (exp).} 
        \label{eq:pot_exp}
\end{align}
The factors in the exponential function were chosen such that, besides at $q=0$, all potentials have  
the same value at $\tilde{q}_c = 3 q_c$; at this momentum they have decayed to $\approx 10^{-4}$ 
of the $q=0$ value and we can safely set the potentials to $0$ for $q > \tilde q_{\rm c}$. 
Considering larger  $\tilde{q}_c$ we have verified that this does indeed not affect our results. 
For small 
momenta the potentials go as $1-g_{4,\nu}(q)/ g_{4,\nu}  \sim |q/q_{\rm c}|^p $, with $p=\infty$ for the box,
$p=4$ for the `p=4'-potential, $p=2$ for the Gaussian potential, and $p=1$ for the exponential potential. 
The exponent $p$ is a measure for the flatness of the potential at $q=0$; see also Ref.~\onlinecite{Meden99}.

For definiteness in our calculations we have always chosen $g_{4,c}/(\pi v_{\rm F}) = 1/2 =-g_{4,s} 
/ ( \pi v_{\rm F})$ without affecting our conclusions. In this case $v_c > v_s$. The measure 
for the system size is $n_{\rm c}$. To numerically compute the recursively defined coefficients 
within reasonable time we choose $n_{\rm c} = 5 \cdot 10^4$. The broadening $\chi$ is chosen such that 
in the broadened spectral function no effects of the individual $\delta$-peaks are 
visible. For the given $n_{\rm c}$      
we take  $\chi / (v_{\rm F} q_{\rm c}) = 5 \cdot 10^{-5}$. 

\begin{figure}[h!]
\includegraphics[width=1.\linewidth,clip]{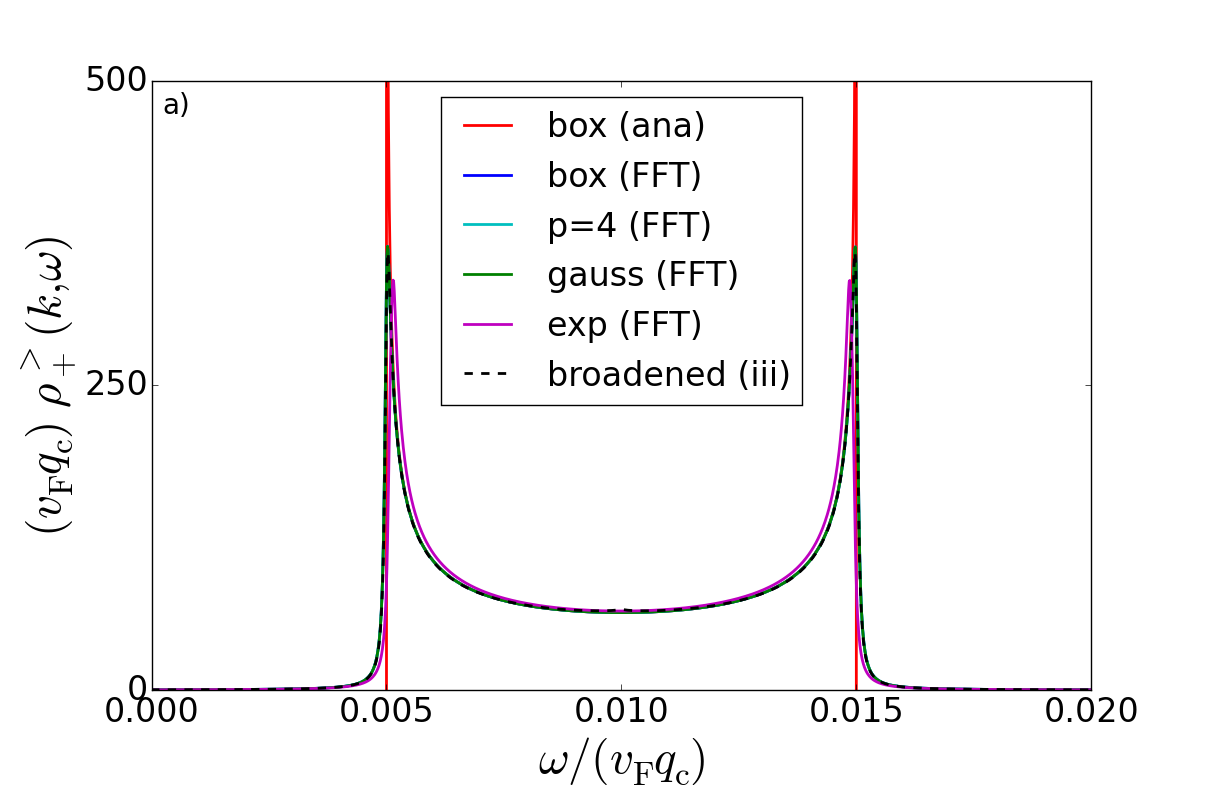}
\includegraphics[width=1.\linewidth,clip]{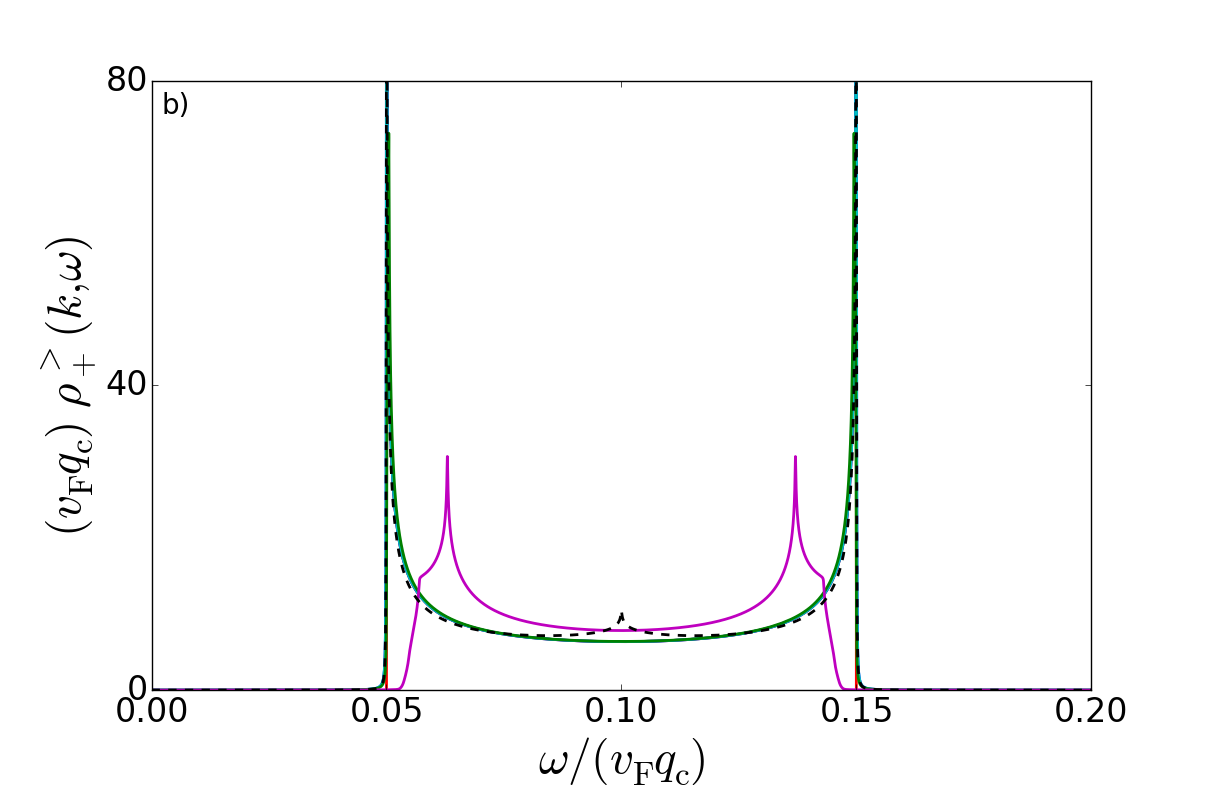}
\includegraphics[width=1.\linewidth,clip]{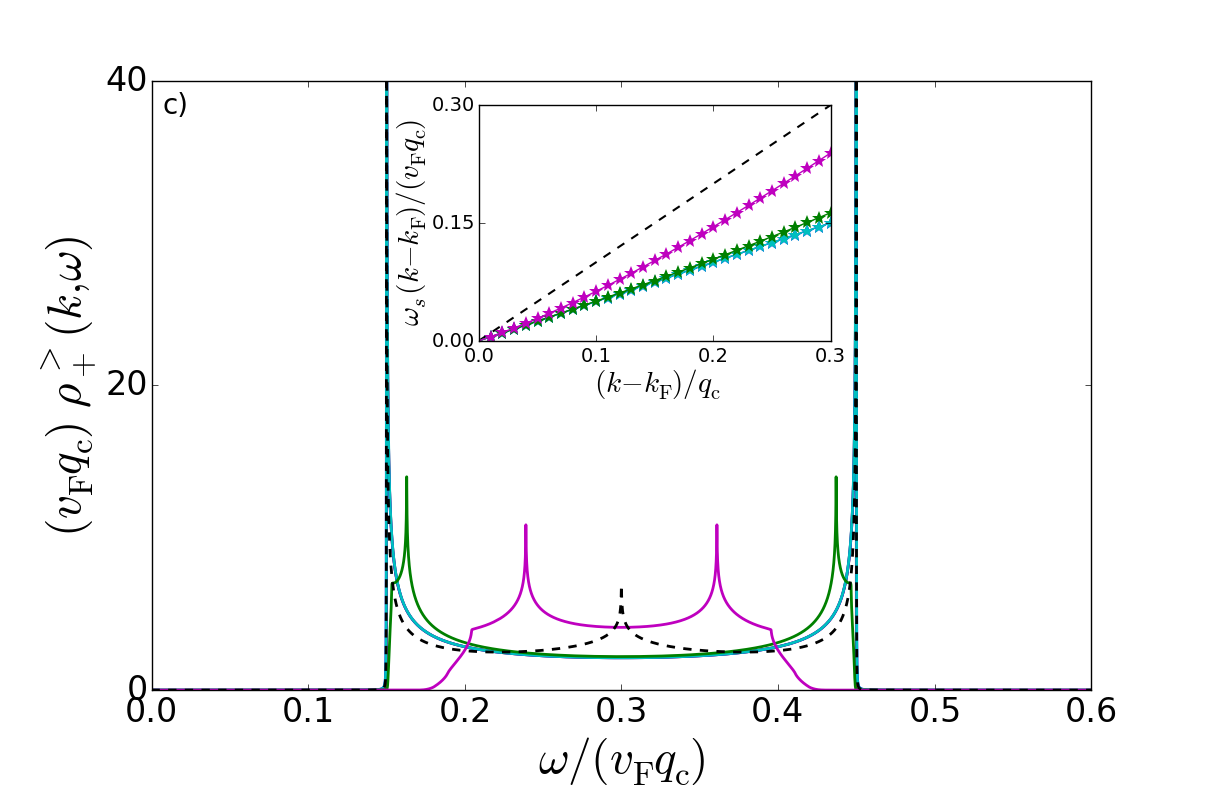}
\caption{(Color online) Spectral function of the spinful $g_4$-model as a function 
of energy for a) $(k-k_{\rm F})/q_{\rm c}=0.01$, b)  $(k-k_{\rm F})/q_{\rm c}=0.1$, and c)   
$(k-k_{\rm F})/q_{\rm c}=0.3$. Spectra for the different potentials are shown. In addition to 
the FFT data with broadening $\chi / (v_{\rm F} q_{\rm c}) = 5 \cdot 10^{-5}$ 
we show unbroadened results for the box potential [`box (ana)'] and the broadened 
results of the ad hoc procedure (iii) [`broadened (iii)']. The other parameters 
are $g_{4,c}/(\pi v_{\rm F}) = 1/2 =-g_{4,s} 
/ ( \pi v_{\rm F})$ and   $n_{\rm c} = 5 \cdot 10^4$. 
The curves are partly hidden by others (see the text). 
The inset of c) shows the 
position of the spin peak extracted from the data (stars) in comparison 
to the collective spin mode dispersion $\omega_s(k-k_{\rm F})$ (lines). The dashed
line displays the unrenormalized dispersion $v_{\rm F}(k-k_{\rm F})$.      
}
\label{fig1}
\end{figure}

In Figs.~\ref{fig1}a) to \ref{fig1}c) we show  $\rho^>_{+} (k, \omega)$ for three different 
small $(k-k_{\rm F})/q_{\rm c}$. The curves labeled as `box (ana)' are the results for the box 
potential obtained from Eq.~\eqref{eq:g4spinrhoex} without convoluting it with a 
Lorentzian. The weights of the $\delta$-peaks are divided by the level spacing (which for 
the given parameters is $1/n_{\rm c}$) and are connected to form a smooth curve. All other results 
(up to `broadened (iii)'; see below)  were obtained from Eq.~\eqref{eq:g4spinG(k,t)} 
multiplied with an exponentially decaying function $\exp\{-\chi |t| \}$ and transformed with 
a FFT; for comparison we also show the broadened spectra for the box potential [box (FFT)]. 

As can be seen 
from Fig.~\ref{fig1}a), for $k$ very close to $k_{\rm F}$ 
all curves are nearly indistinguishable. The broadening of the spectra is visible by the weight 
`leaking out' for $\omega < v_s [k-k_{\rm F}] $ and $\omega > v_c [k-k_{\rm F}]$.  However, 
already for $(k-k_{\rm F})/q_{\rm c} = 0.1$ [Fig.~\ref{fig1}b)], the spectral function calculated 
with an exponentially decaying potential shows pronounced differences to the other ones. 
For $(k-k_F)/q_c = 0.3$, also the curve of the Gaussian potential deviates from the one of the box 
potential, see Fig.~\ref{fig1}c). The spectrum of the `very flat' p=4 potential still lies
on top of the one obtained for the box potential; differences only appear at even larger 
$k-k_{\rm F}$ (not shown). 
Obviously, the smaller the above introduced index $p$, that is the less `flat' the potential is 
at $q=0$, the faster the line shape of the spectra starts to deviate from the one of the box potential
when $k- k_{\rm F}$ increases.

The maxima of $\rho^>_+(k,\omega)$ are located at 
$\omega_{\nu}(k-k_{\rm F})$, instead of at $v_\nu [k-k_{\rm F}]  $; for $|k-k_{\rm F}|< q_{\rm c}$ 
both positions are equal for the box potential. 
This is shown in the inset of Fig.~\ref{fig1}c), where the spin dispersion relation for the 
different potentials (full lines) is compared to the numerically determined maxima of the spectra 
(stars). The agreement is very good; the charge peak behaves similarly. In the limit 
of small $k-k_{\rm F}$ and large $p$ the difference between  $\omega_{\nu}(k-k_{\rm F})$ and 
$ v_\nu [k-k_{\rm F}]$ is negligible.  

In the literature the third ad hoc regularization procedure leading to 
$[G_{+}^{>}]_{\rm (iii)} (x,t)$  Eq.~(\ref{gapprox3}) is considered `the best' one\cite{Voit93b,Voit95}
as it leads to a $[\rho^>_+]_{\rm (iii)}(k,\omega)$ which fulfills exact sum rules.\cite{Suzumura80,Schoenhammer93b} 
For $\gamma_c=\gamma_s=0$ a closed analytical expression of the double Fourier transform
can be given; see Eqs.~(3.20) and (3.21) of Ref.~\onlinecite{Voit93b}.  
We therefore added a graph of this analytical result convoluted with a Lorentzian of width $\chi$
as the dashed lines in Figs.~\ref{fig1}a) to \ref{fig1}c) (with $1/\Lambda \to q_{\rm c}$). 
As mentioned above these curves as well as the exact function for the box potential show single-sided 
(threshold) square-root singularities at $v_{\nu} [k-k_{\rm F}]$ (for a more detailed analysis on this see below).     
The approximate spectrum $[\rho^>_+]_{\rm (iii)}(k,\omega)$ is characterized by an
additional feature at $\omega= v_{\rm F} [k-k_{\rm F}]$ (barely visible in Fig.~\ref{fig1}a)
for $(k-k_{\rm F})/q_{\rm c}=0.01$); as analyzed in Ref.~\onlinecite{Voit93b} a logarithmic
divergence appears at this energy. It results from the additional factor  
$(x- v_{\rm F}  t + i \Lambda)/(x- v_{\rm F}  t +i0^+)$ of $[G_{+}^{>}]_{\rm (iii)} (x,t)$ 
Eq.~(\ref{gapprox3}). Obviously, the exact spectral function of the $g_4$-model does not 
display this feature for any of the considered potentials. For the box potential, which besides 
the missing peak at $\omega= v_{\rm F} [k-k_{\rm F}]$ shows the same spectral characteristics
as found in the ad hoc procedure, this can even be understood analytically: as seen 
in Eq.~(\ref{eq:g4spinrhoex}) for $k-k_{\rm F} < q_{\rm c}$ the information on $v_{\rm F}$ 
drops completely out. The logarithmic divergence 
of $[\rho^>_+]_{\rm (iii)}(k,\omega)$   at $\omega= v_{\rm F} [k-k_{\rm F}]$ is thus an {\it artifact 
of the ad hoc regularization.} As discussed very recently, this logarithmic divergence 
for $k-k_F \lessapprox q_{\rm c}$ turns into a 
power-law one if the spinful  ($g_2$-$g_4$) TLM is treated within the ad hoc procedure 
(iii).\cite{Maebashi14} Also this feature is an artifact of the ad hoc procedure. 
With Ref.~\onlinecite{Maebashi14} in mind we emphasize that this does not exclude 
that for $k-k_{\rm F} \gg q_{\rm c}$ all the spectral weight is located around $\omega =  
v_{\rm F} (k-k_{\rm F})$, which, in fact, is generically 
the case. This was discussed for the box potential in Ref.~\onlinecite{Schoenhammer93} and for 
general interactions in Ref.~\onlinecite{Schoenhammer93b}. We here do not investigate this any
further as we are exclusively interested in the spectra at small $|k-k_{\rm F}|$.

In order to investigate whether or not the spectra at fixed $0< k-k_{\rm F} \ll q_{\rm c}$ and 
for $\omega$ close to the maxima at $\omega_{\rm max}$ 
show power-law scaling we do not 
simply want to rely on the quality of power-law fits to the broadened data. Instead we take 
the logarithmic derivative 
\begin{equation}
\mbox{logder}(\omega)=\frac{ d \ln \left[ \rho^>_{+} (k, \omega) \right] }{d \ln | \omega- \omega_{\rm max} | }, 
\label{eq:logder}
\end{equation} 
of our data with $\omega_{\rm max}$ equal to the corresponding peak positions. It is numerically approximated by 
centered differences and provides a very sensitive approach in the search for power laws. 
If for $\omega \to \omega_{\rm max}$, $\mbox{logder}(\omega)$ approaches a constant $\xi$ 
the spectral function shows power-law scaling with exponent $\xi$ close to $\omega_{\rm max}$ (according 
to the definition given in the introduction).      

\begin{figure}[th!]
\includegraphics[width=1\linewidth,clip]{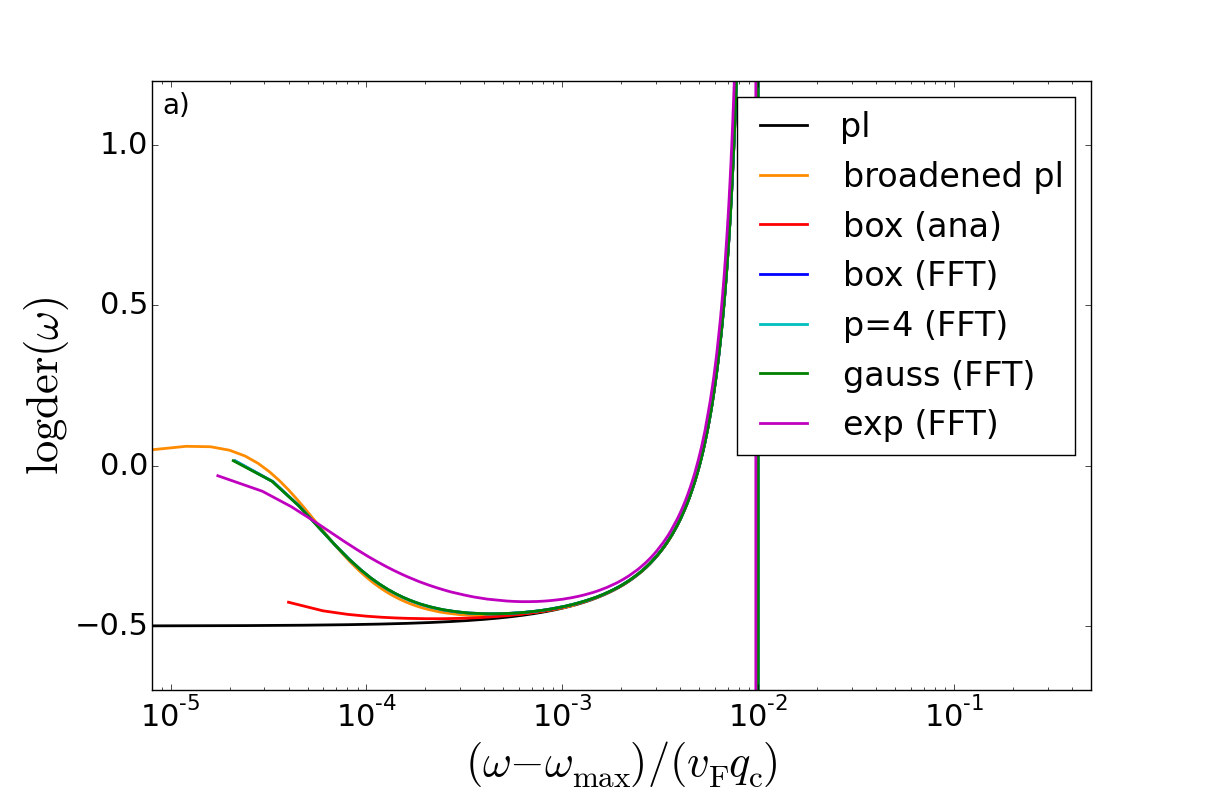}
\includegraphics[width=1\linewidth,clip]{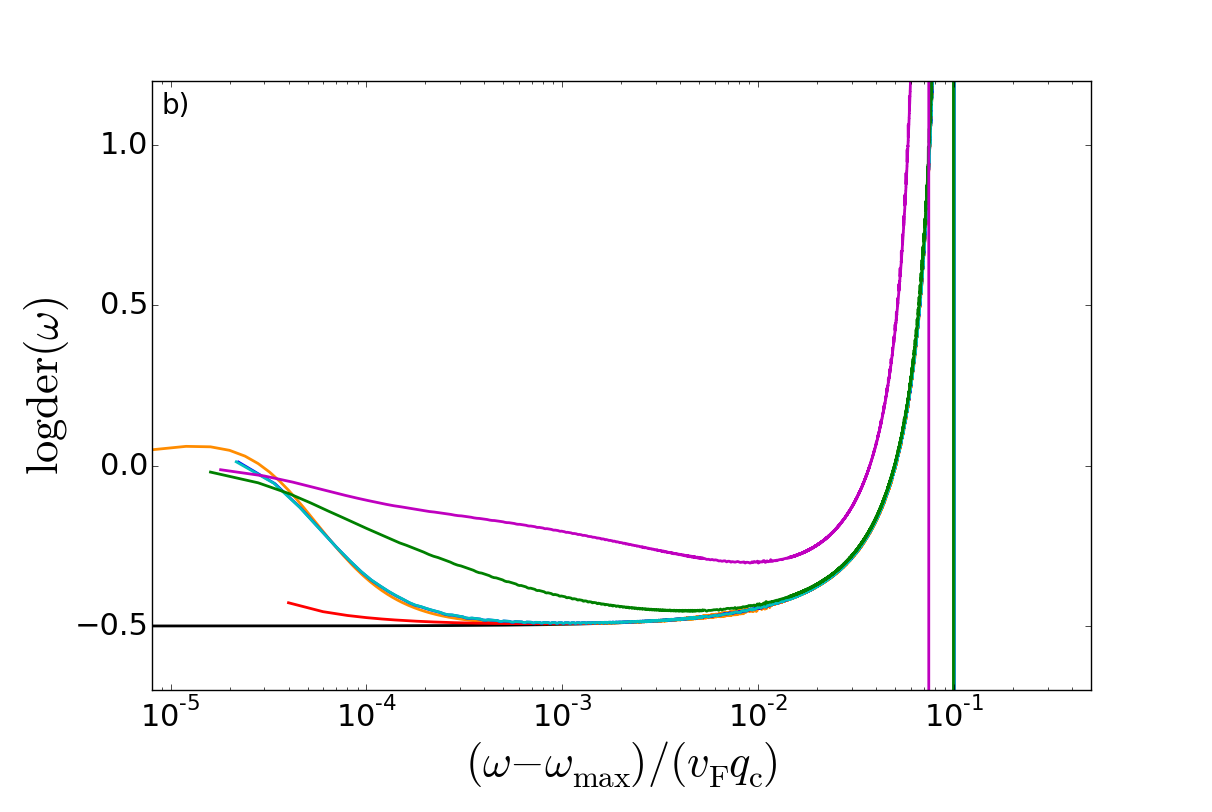}
\caption{(Color online) Logarithmic derivative Eq.~(\ref{eq:logder}) of the 
spectral function of the spinful $g_4$-model close to the spin peak 
for a) $(k-k_{\rm F})/q_{\rm c}=0.01$ and b)  $(k-k_{\rm F})/q_{\rm c}=0.1$. In addition to 
the broadened FFT data for the different potentials and the 
unbroadened results for the box potential [`box (ana)']
we show broadened [`broadened pl'] as well as unbroadened  
[`pl'] data for the product of single-sided square-root singularities Eq.~(\ref{eq:g4powlaw}).
The parameters are as in Fig.~\ref{fig1}.
}
\label{fig2}
\end{figure}

In Figs.~\ref{fig2}a) [$(k-k_{\rm F})/q_{\rm c}=0.01$] and \ref{fig2}b)  [$(k-k_{\rm F})/q_{\rm c}=0.1$] 
we show $\mbox{logder}(\omega)$ close to the spin peak for the different
two-particle potentials. For symmetry reasons the behavior close to the charge peak is the same.
Instead of an ad hoc regularized spectral function -- which is spoiled by the spurious peak at 
$\omega =  v_{\rm F} (k-k_{\rm F})$ --  we this time present the logarithmic derivative  
of the simple normalized product of two single-sided square-root singularities 
\begin{equation}
	[\rho^>_+]_{\mathrm{pl}} (k,\omega) = \frac{1}{\pi} \, \frac{ \Theta 
\left( \omega - v_s [k-k_F] \right) }{ \left( \omega - v_s [k-k_F] \right)^{1/2} } \, 
\frac{ \Theta \left( v_c [k-k_F] - \omega \right) }{ \left(  v_c [k-k_F] -  \omega \right)^{1/2} },
	\label{eq:g4powlaw}
\end{equation}
for reference. It is indicated by `pl'. In addition we present the logarithmic derivative of this 
expression convoluted with a Lorentzian of width  $\chi$, indicated by `broadened pl'. 
The unbroadened data for the box potential -- for which we analytically know that for $L \to  \infty$  
a square root singularity at $v_s [k-k_{\rm F}]$ exists [see Eq.~(\ref{wurz})] --  very nicely follow the `pl' curve 
down to $\omega - \omega_{\rm max} \approx 4 \cdot 10^{-4} v_{\rm F} q_{\rm c}$.
 At this energy {\it finite size corrections} 
destroy the power-law scaling. This can be verified by considering different $n_{\rm c}$, that is 
different system sizes. The `broadened pl' curve starts to deviate from the unbroadened one 
at $\omega - \omega_{\rm max} \approx 6 \cdot 10^{-4} v_{\rm F} q_{\rm c} $.
From this we conclude that for the chosen $n_{\rm c}$ 
and $\chi$ the broadening $\chi$ cuts off the power-law scaling at slightly larger energies
than the system size.
This is consistent with the observation that the broadened spectrum of the box potential [`box (FFT)'] 
almost perfectly follows the  `broadened pl' curve down to the much smaller scale 
$\omega - \omega_{\rm max} \approx 6 \cdot 10^{-5} v_{\rm F} q_{\rm c}$. This gives us confidence that the FFT data for 
the other potentials are unspoiled by both finite size and broadening effects down to  
$\omega - \omega_{\rm max} \approx 6 \cdot 10^{-4} v_{\rm F} q_{\rm c}$. At the `high energy' end possible power law 
scaling close to the spin peak is cutoff by the charge peak. The latter is the origin of the 
dominant feature at $\omega-\omega_{\rm max} \approx 10^{-2}  v_{\rm F} q_{\rm c}$ and 
$\omega-\omega_{\rm max} \approx 10^{-1}  v_{\rm F} q_{\rm c}$, respectively.      

While for $(k-k_{\rm F})/q_{\rm c}=0.01$ the data of the `p=4' and the `gauss' potential 
lie perfectly on top of the `box (FFT)' data, and one is tempted to conclude that they are consistent with a 
square-root singularity at $\omega_{\rm max}$, the data for the exponential potential clearly differ 
and are inconsistent with such behavior. For $(k-k_{\rm F})/q_{\rm c}=0.1$ in addition 
the data for the Gaussian potential are incompatible with this type of singularity. For even 
larger $k-k_{\rm F}$ (but still smaller than $q_{\rm c}$; not shown) also the data for the 
`p=4' potential no longer follow the ones of the box potential. 

The most consistent interpretation of our results is that for any potential which is not 
`infinitely flat' at $q=0$, that is 
if $p<\infty$, the spectral function at fixed $k-k_{\rm F} >0$ does strictly speaking not show 
power-law scaling close to $\omega_{\rm max}$. The less `flat' the potential is, that
is, the smaller $p$, the faster this becomes apparent as $k-k_{\rm F}$ is increased. The 
square-root singularities found for the box potential (and for the ad hoc procedures) are cut 
off by the curvature of the potential close to $q=0$. 
They can thus not be considered as universal features of the spinful $g_4$-model. 

We note that the data for $p<\infty$ are not only inconsistent with power-law scaling 
when taking $\omega_{\rm max}$ as the point of reference. We studied the behavior relative to 
other distinguished energies 
(e.g. $v_\nu [k-k_{\rm F}]$ and the -- due to the broadening -- apparent thresholds). For 
none of these we find behavior which is consistent with power laws.       

Despite this lack of power-law behavior, for all interaction potentials studied 
the exact spectral function of the $g_4$-model is still characterized by spin and 
charge peaks.
             
\subsection{Spectra of the spinless $g_2$-$g_4$-model} 
\label{sect_g2_g4}

Employing the formulas given in Sects.~\ref{nospinbox} and \ref{nospinarb} we can compute 
the exact spectral function of the spinless TLM -- the spinless $g_2$-$g_4$-model -- for
different forms of the potential. In the last section we saw that for the small 
$k-k_{\rm F}$ we are interested in, the spectra for the `p=4' potential Eq.~(\ref{p4pot}) 
barely differ from the ones obtained for the box potential. The same holds for the full 
spinless model and in this section we focus on the box potential, the Gaussian one Eq.~(\ref{eq:pot_gaus}) 
as well as the exponential potential Eq.~(\ref{eq:pot_exp}). To prevent an inflation of 
cases and parameters we consider the physically reasonable situation of equal inter- and 
intra-branch scattering $g_2(q) = g_4(q) = g(q)$. We choose $g(0)$ such that $v_c=2 v_{\rm F}$ and $\gamma_c(0)=1/8$.

To compute $\rho^>_+(k,\omega)$ via $G^>_+(k,t)$ Eq.~(\ref{eq:g2g4spinless_G_coeff}) 
and FFT for general potentials or via Eq.~(\ref{eq:g2g4rho_boxex}) for the box potential  
we have to recursively compute more coefficients and perform additional sums in comparison 
to what was necessary in the spinless $g_4$-model. This increases 
the numerical resources required and we thus have to consider smaller system sizes compared 
to the latter; we choose $n_{\rm c} = 2 \cdot 10^4$. Furthermore, at fixed $n_{\rm c}$ 
the energy level spacing of the spinless $g_2$-$g_4$-model is larger than that of the spinful 
$g_4$-model. To obtain smooth curves we thus have to increase the broadening $\chi$. We take 
$\chi/(v_{\rm F} q_{\rm c}) = 10^{-3}$.       

\begin{figure}[h!]
\includegraphics[width=1.07\linewidth,clip]{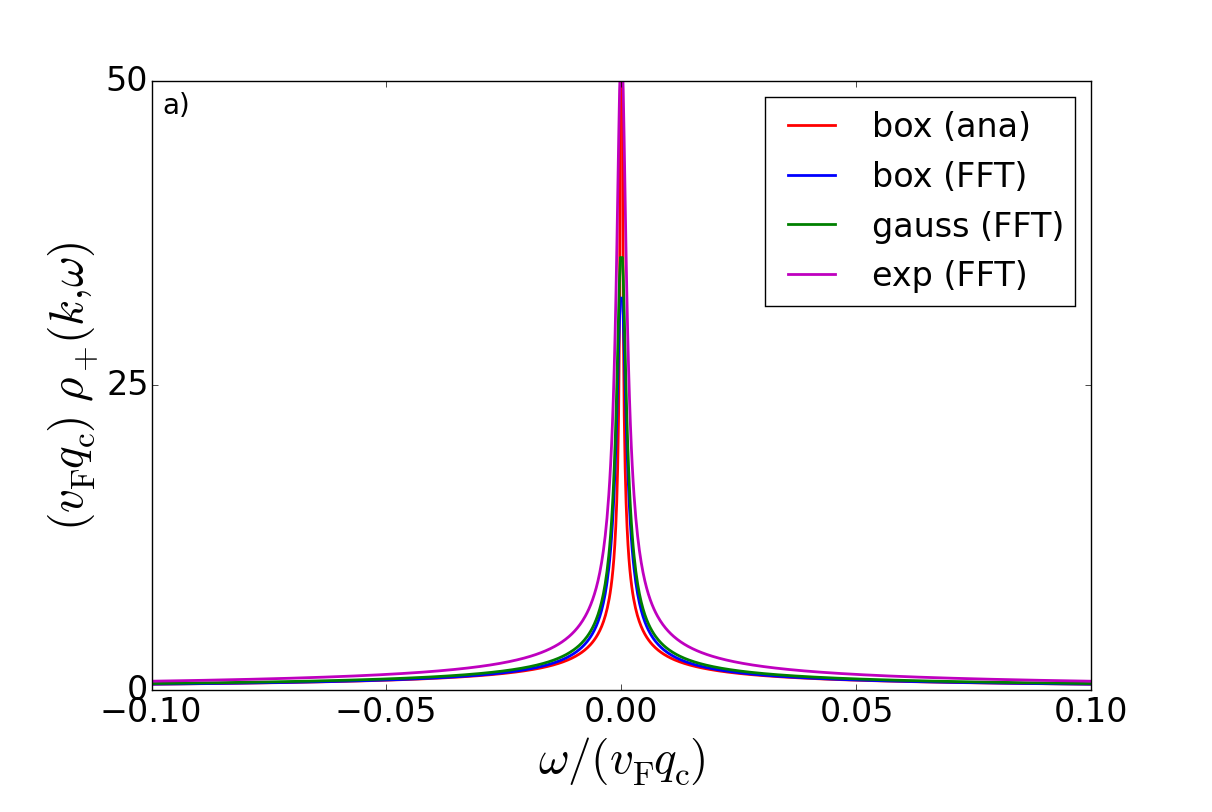}
\includegraphics[width=1.07\linewidth,clip]{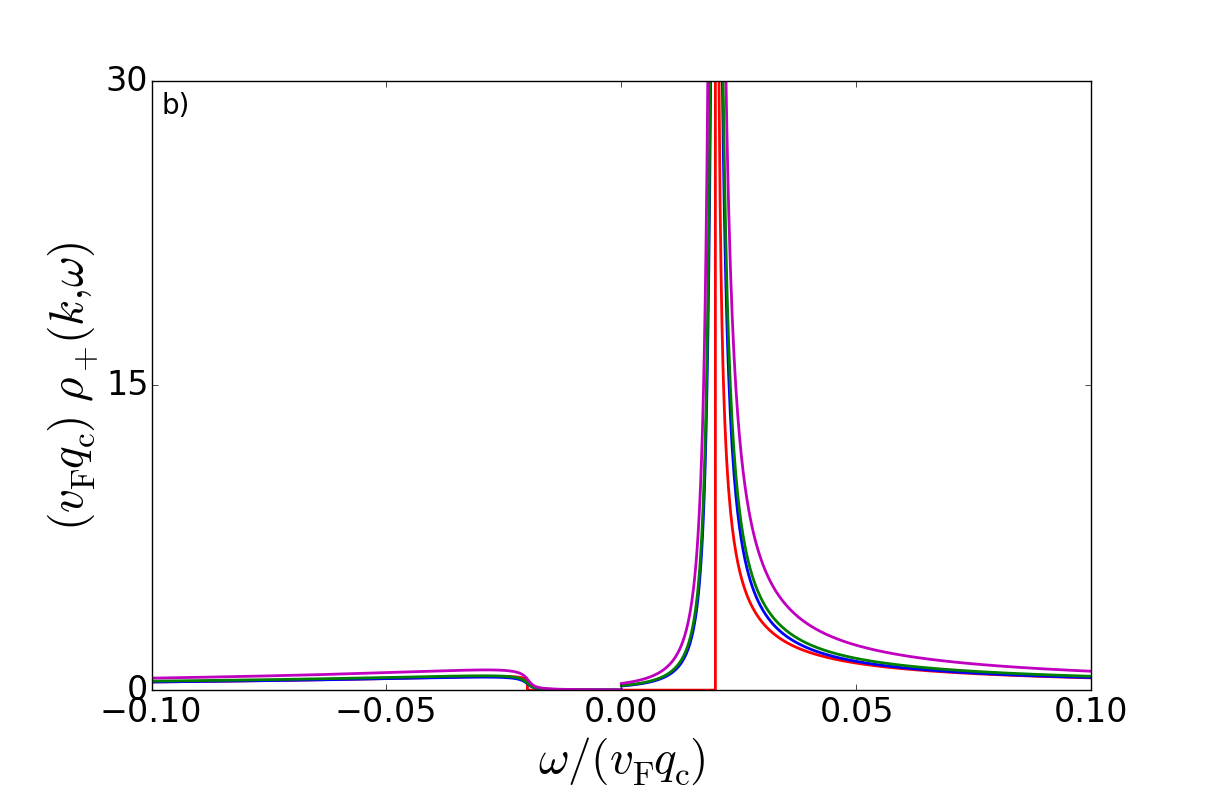}
\includegraphics[width=1.07\linewidth,clip]{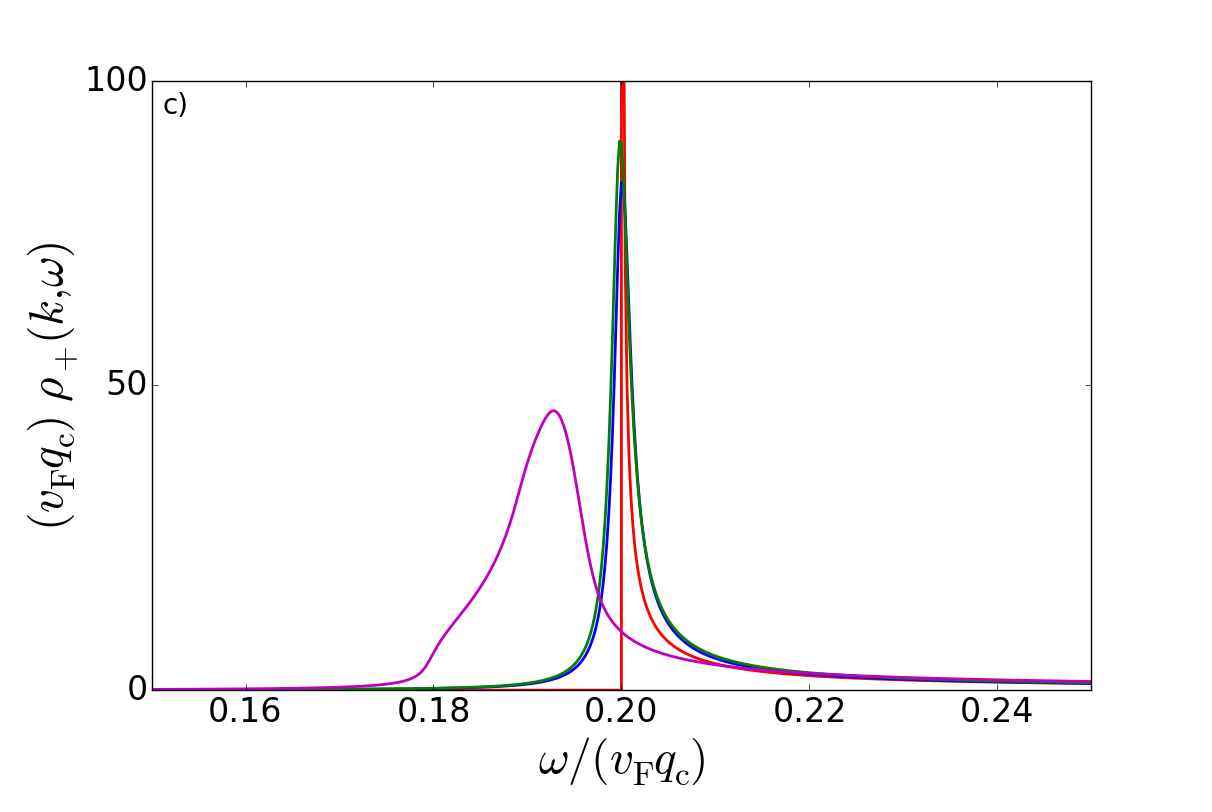}
\caption{(Color online) Total spectral function of the $g_2$-$g_4$-model as a function 
of energy for a) $(k-k_{\rm F})/q_{\rm c}=0$, b)  $(k-k_{\rm F})/q_{\rm c}=0.01$, and c)   
$(k-k_{\rm F})/q_{\rm c}=0.1$. Spectra for the different potentials are shown. In addition to 
the FFT data with broadening $\chi / (v_{\rm F} q_{\rm c}) = 10^{-3}$ 
we show unbroadened results for the box potential [`box (ana)']. The 
interaction at $q=0$ is chosen such that $v_c = 2 v_{\rm F}$ and $\gamma_c(0)=1/8$. The 
system size is set by $n_{\rm c} = 2 \cdot 10^4$. 
}
\label{fig3}
\end{figure}

In Figs.~\ref{fig3}a) to \ref{fig3}c) we present the {\it total spectral function} 
$\rho_+(k,\omega) = \rho^>_{+} (k, \omega) + \rho^<_{+} (k, \omega)$ for $(k-k_{\rm F})/q_{\rm c}=0$, $0.01$, 
and $0.1$. We switched to this as it simultaneously shows the photoemission as well 
as inverse photoemission part of the spectrum. For increasing $k-k_{\rm F }$ the photoemission 
part looses weight quickly. Thus in Fig.~\ref{fig3}c) with $(k-k_{\rm F})/q_{\rm c}=0.1$ we only 
present a zoom-in of the inverse photoemission part. In addition to the broadened functions for 
the box, Gaussian, and exponential potentials obtained by FFT we show the unbroadened one of the 
box potential [see Eq.~(\ref{eq:g2g4rho_boxex})]. As above, to obtain the latter the weights of 
the $\delta$-peaks were divided by the level spacing (which for the given parameters is $4/n_{\rm c}$) 
and are connected to form a smooth curve. 

As discussed in Sect.~\ref{nospinbox} for $0< k-k_{\rm F} < q_{\rm c}$ and the box potential 
$\rho_+(k,\omega)$ shows threshold power-law nonanalyticities at $\pm v_c [k-k_{\rm F}]$ with 
exponents $\gamma_c-1$ (for $\omega>0$) and $\gamma_c$ (for $\omega <0$). In `box (FFT)' these 
are broadened. Similarly to the spinful $g_4$-model we observe that the smaller $p$ the faster 
the line shape starts to deviate from the one of the box potential when increasing $k-k_{\rm F}$. 
For   $(k-k_{\rm F})/q_{\rm c}=0.1$ [see Fig.~\ref{fig3}c)] and the exponential potential with 
$p=1$ this already leads to a strongly modified  distribution of the spectral weight. The 
deformed line shape can be understood in due detail when comparing it to the spectral function 
of the {\it spinless} $g_4$-model.\cite{Markhof15} This detailed analysis is beyond the 
scope of the present paper. 

\begin{figure}[h!]
\includegraphics[width=1.\linewidth,clip]{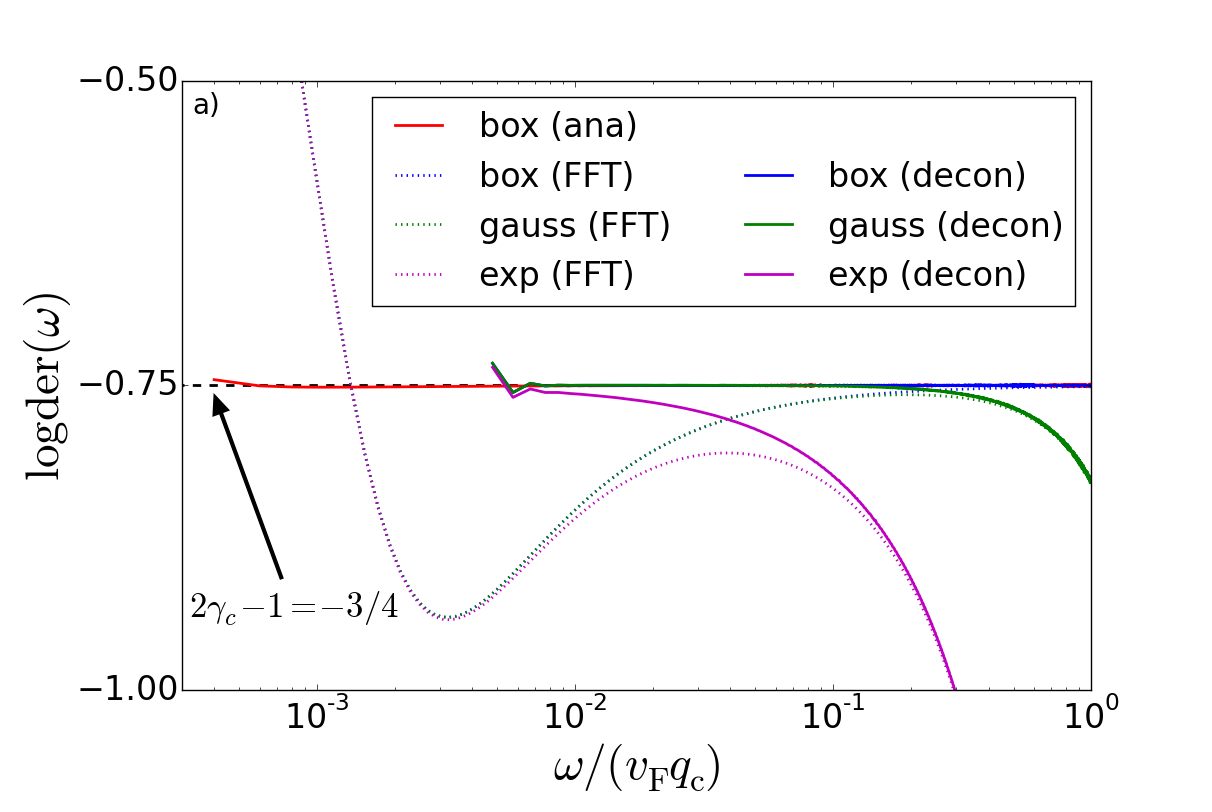}
\includegraphics[width=1.\linewidth,clip]{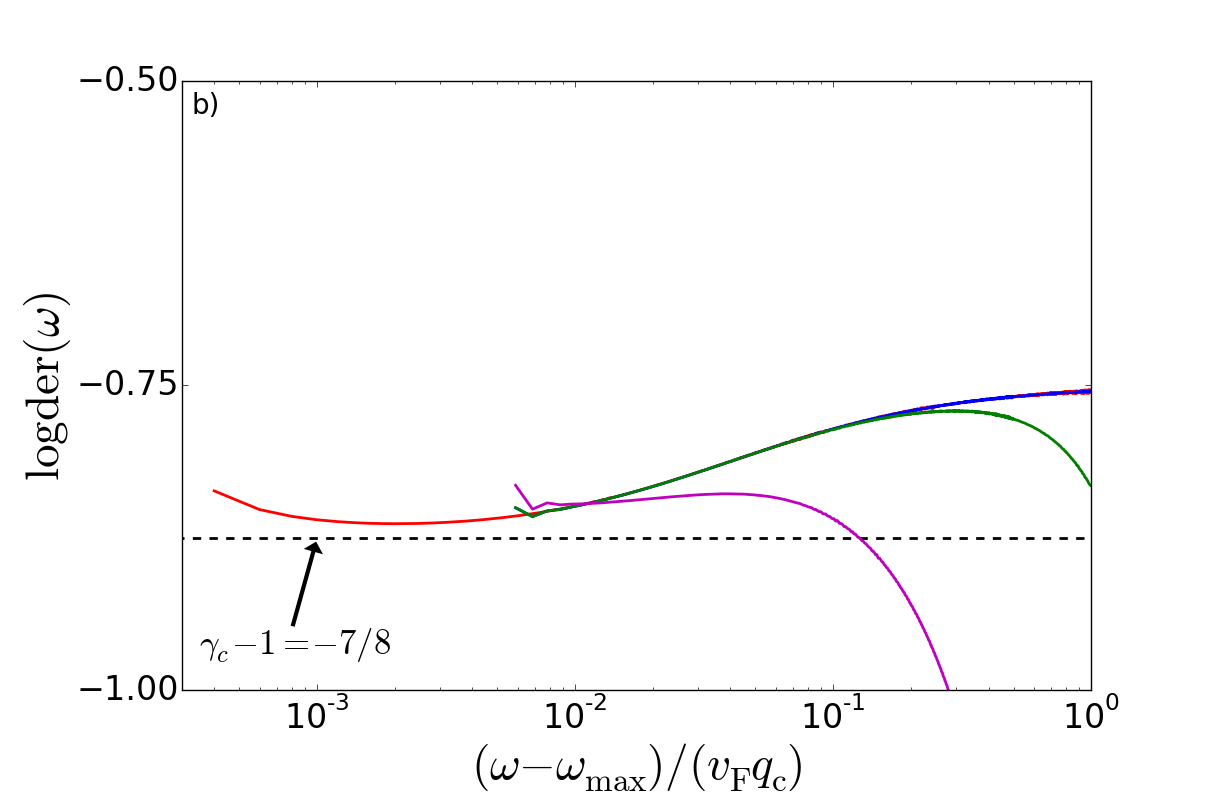}
\includegraphics[width=1.\linewidth,clip]{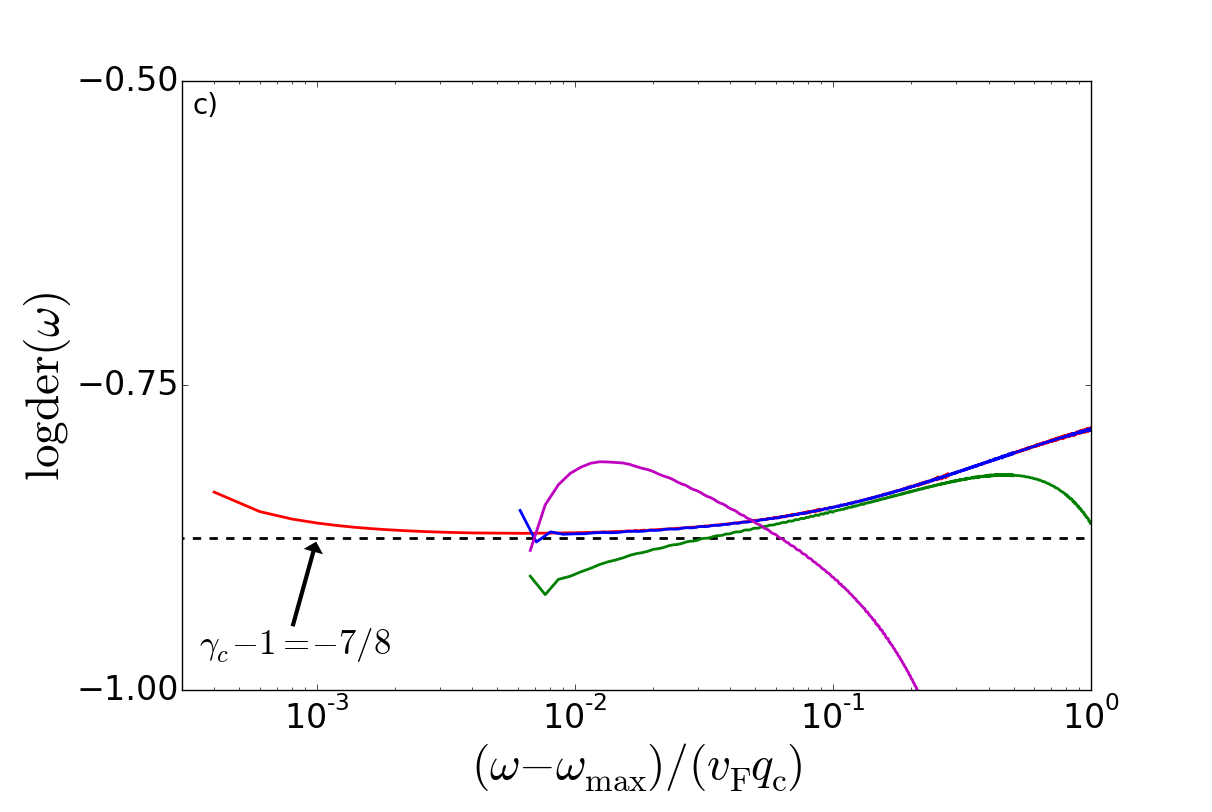}
\caption{(Color online) 
Logarithmic derivative Eq.~(\ref{eq:logder}) of the 
spectral function of the spinless $g_2$-$g_4$-model close to the 
inverse photoemission peak 
for a) $(k-k_{\rm F})/q_{\rm c}=0$ and b)  $(k-k_{\rm F})/q_{\rm c}=0.01$, and 
c)  $(k-k_{\rm F})/q_{\rm c}=0.1$.
The full lines labeled by `box (ana)' show the unbroadened results of the box
potential. The dotted lines in a) result from broadened spectra. The other 
full lines are obtained from the numerical spectra after a deconvolution 
(see the text). The dashed lines indicate the expected exponent.   
The parameters are as in Fig.~\ref{fig3}.
}
\label{fig4}
\end{figure}

From Ref.~\onlinecite{Meden99} and the RG irrelevance 
of the momentum dependence of the two-particle potential we expect that for all potentials 
$\rho_+(k_{\rm F},\omega) \sim |\omega|^{2 \gamma_c -1}$. On first glance the data of Fig.~\ref{fig3}a)                   
appear to be consistent with this behavior, however, a more thorough analysis is required. 
In  Fig.~\ref{fig4}a) we plot the logarithmic derivative Eq.~(\ref{eq:logder}) (with $\omega_{\rm max}=0$) 
of the $\omega>0$ broadened FFT data of Fig.~\ref{fig3}a) as dotted lines. While the logarithmic 
derivative of the unbroadened `box (ana)' data (solid line) nicely shows a plateau at the 
expected exponent $2 \gamma_c -1 = -3/4$ (for the given parameters), which is only spoiled at 
very small $\omega/(v_{\rm F} q_{\rm c}) \approx 10^{-3}$ due to finite size effects, the dotted 
curves do not seem to support power-law scaling of  $\rho_+(k_{\rm F},\omega)$. This also holds 
for the {\it broadened} `box (FFT)' data which establishes that the {\it broadening destroys the 
power law even for} $\omega \gg \chi$.  

To further analyze this surprising finding in a `controlled' setup we took the function $f(x) = x^{-3/4}$ 
and convoluted it with a Lorentzian of width $10^{-3}$. We indeed found that for $x \in [10^{-2},10^{-1}]$ 
the logarithmic derivative of the resulting function shows a behavior quite similar to the one of the 
dotted lines in Fig.~\ref{fig4}a); in particular, it bends away from the expected plateau towards smaller
values. For larger exponents, let's say -1/2 as in Figs.~\ref{fig2}a) and b), this does not happen. One is 
thus tempted to increase the interaction strength and thus $2 \gamma_c -1$. However, larger interactions
also imply larger energy level spacings of the $\delta$-peaks and thus require larger $\chi$, which cuts off 
the power-law scaling at larger energies.                 

As a possible way out we performed a {\it deconvolution} of our numerically obtained spectral function along 
the lines of Refs.~\onlinecite{Gebhard03} and \onlinecite{Raas05}. The resolution was chosen to be 
approximately equal to the broadening $\chi$. The deconvolution of numerical data is of course an 
ill-posed problem and the corresponding spectral function can e.g. become negative.\cite{Raas05} 
We were able to perform a stable deconvolution down to $(\omega-\omega_{\rm max})/(v_{\rm F} q_{\rm c}) 
\approx 10^{-2}$. In fact, the resulting spectra are {\it sufficiently smooth to perform 
logarithmic derivatives.}  These are shown as solid lines in Fig.~\ref{fig4}a). The oscillatory 
behavior at the lower end signals the onset of an instability of the deconvolution. The deconvoluted 
data for the box potential lie exactly on top of the ones obtained from the unbroadened spectral 
function `box (ana)'. This gives us confidence that also for the other potentials the deconvolution 
can be trusted. This is further supported by the observation that the data for the Gaussian potential 
now clearly support power-law scaling with exponent $2 \gamma_c -1$. For the exponential potential the energy 
scale cutting off the power-law scaling at the {\it `high energy' side} appears to be of the order 
$10^{-2}$ and thus smaller compared to the one of the box (of the order $1$)  and the Gaussian potential 
(of the order $10^{-1}$). Therefore no clear plateau is reached for the accessible energies. 
However, the data appear to saturate at the expected value $2 \gamma_{c}-1$. 
  
Along the same lines we next investigate whether or not for $k-k_{\rm F} > 0$ a power law is found 
close to the peak  on the inverse photoemission side. In Figs.~\ref{fig4}b) 
[$(k-k_{\rm F})/q_{\rm c}=0.01$] and c) [$(k-k_{\rm F})/q_{\rm c}=0.1$] we 
present the logarithmic derivative of the deconvoluted spectra together with `box (ana)' data. 
The logarithmic derivative of the original spectra behave similar to the $k=k_{\rm F}$ case and are 
thus not shown.  As expected the `box (ana)' data are consistent with power-law scaling with exponent 
$\gamma_c-1 = -7/8$  (for the given parameters). The power law is cut off at   $(\omega-\omega_{\rm max})
/(v_{\rm F} q_{\rm c}) \approx 10^{-3}$ due to finite size effects. The deconvoluted box potential data 
fall again exactly on top of  the `box (ana)' ones indicating that the deconvolution is stable. 
For $(k-k_{\rm F})/q_{\rm c}=0.01$ the data for the Gaussian potential are for sufficiently small 
$\omega-\omega_{\rm max}$ on top of the box potential ones [see Fig.~\ref{fig4}b)]. 
This does no longer hold for   $(k-k_{\rm F})/q_{\rm c}=0.1$  [see Fig.~\ref{fig4}c)] 
for which the data of the Gaussian potential are incompatible with power-law scaling. 
For the exponential potential with smaller $p$ this deviation from possible power-law behavior
sets in at already smaller $k-k_{\rm F}$ [see Fig.~\ref{fig4}b)]. 

As for the spinful $g_4$-model
the most consistent interpretation of our results is that for any potential with $p<\infty$ 
and fixed $k-k_{\rm F} \neq 0$ the power-law scaling found for the box potential is destroyed by 
the curvature of the potential at $q=0$; power laws are thus nongeneric. 

We note that this does not only hold when taking $\omega_{\rm max}$ as the 
point of reference. We studied the behavior relative to other distinguished energies 
(e.g. $v_c [k-k_{\rm F}]$ and the -- due to the broadening -- apparent threshold). For none of these 
points we find behavior which is consistent with power laws. Similarly also the behavior close 
to the threshold on the photoemission side are incompatible with power-law scaling if $p<\infty$.           
 
These results imply that the spectral functions resulting from the different ad hoc 
regularizations (not shown here), which are all characterized by the threshold power laws 
Eqs.~(\ref{power1}) and (\ref{power2}), are nongeneric.   
 
\section{Discussion}
\label{sect_discussion}

Our above results for the exact spectral function can be summarized as follows:
\begin{enumerate}
\item For $k-k_{\rm F} \neq 0$ and generic two-particle potentials which are not `infinitely flat'
at momentum transfer $q=0$ the spectral function does not show power-law scaling close to any 
of the distinguished energies. Power-law behavior is generically only found if all energy scales are sent
to zero, e.g. in $\rho^>(k_{\rm F},\omega)$ for $\omega \to 0$ (and if $g_2(0)$ is finite).   
\item The ad hoc regularized spectra which are commonly studied show finite $k-k_{\rm F}$ power laws 
and can thus not be considered as generic. The origin of this nongeneric behavior is the linearization 
of the spin and charge dispersion relations.
The ad hoc spectra are plagued by  additional artifacts (see Sect.~\ref{sect_g4}).   
\item The less flat the potential at $q=0$, that is the smaller the introduced index $p$, the 
faster the differences of the spectral line shape compared to the one of the box potential --
and in many respects compared to the one of the ad hoc procedures -- 
becomes apparent when $k-k_{\rm F}$ is increased. For small $p$, e.g. the exponential potential 
with $p=1$, already for $(k-k_{\rm F})/q_{\rm c}=0.1$ major differences are apparent; see Figs.~\ref{fig1}b)
and \ref{fig3}c). For $k-k_{\rm F} \neq 0$ the spectral function is still characterized by the 
dispersing spin and charge peaks. 
\end{enumerate} 
To individually study how the curvature of the potential at zero momentum transfer modifies 
the two interaction effects of spin-charge separation and power laws with 
interaction dependent exponents, we have studied the spinful $g_4$-model and the spinless 
$g_2$-$g_4$-model. Combining the two limiting cases it is obvious that the same conclusions 
can be drawn for the full TLM, that is the spinful $g_2$-$g_4$-model. Explicit results 
for the spectral function of the full TLM are presented in Sect.~\ref{sect_con_exp} (see 
Fig.~\ref{fig5}). 
    
We next discuss the implications of these findings. 

\subsection{Luttinger liquid universality} 
\label{sect_con_LLun}

\subsubsection{General considerations}

The TLM forms the low-energy fixed point model under RG flow of a large class of 1d correlated 
fermion models.\cite{Solyom79} Based on this insight it was suggested that the power laws of 
the ad hoc regularized TLM spectral function $\rho_+^>(k \neq k_{\rm F},\omega)$ should be 
observable in other models from this class.\cite{Voit95} One can doubt this on general grounds 
as $k-k_{\rm F} \neq 0$ sets a finite scale thus breaking quantum critical scale invariance 
and cutting off the RG flow.\cite{Meden99} Our results show {\it explicitly} that power laws in 
$\rho^>(k \neq k_{\rm F},\omega)$ are indeed not part of LL universality; if they cannot be 
found in the generic TLM, they cannot expected to be a universal feature of other models.  

We emphasize that LL universality does {\it not} imply that for a given model from the LL 
universality class (other than the TLM itself) one simply has to choose proper 
coupling functions $g_{i,\kappa}(q)$ -- which in any case would be unknown a priori -- 
and can reproduce details of the spectral function of this model at low energies using 
the TLM. All that is known about the spectral properties of a model from the LL universality 
class is that for $\omega \to 0^+$ the scaling relations  $\rho^>(k_{\rm F},\omega) \sim 
\omega^{\alpha-1}$ and $\rho^>(\omega) \sim \omega^\alpha$ hold. 
 
When computing space-time correlation functions of the TLM other than the single-particle 
Green function momentum integrals of similar type as encountered here appear. To evaluate
these ad hoc procedures including the linearization of the collective spin and 
charge dispersion are commonly employed.\cite{Voit95} One prominent example is the 
density-density correlation function for momentum transfer close to $2 k_{\rm F}$.\cite{Luther74}
As for the single-particle spectral function after regularization the Fourier integrals 
can be performed analytically leading to power-law scaling of the susceptibility with 
interaction dependent exponents close to the characteristic energies 
$\pm v_c [k \pm  2 k_{\rm F}]$. Our considerations imply that also this feature can most 
likely not be considered as being characteristic for LLs in general. The density response 
at {\it vanishing energy} and for $k \to \pm 2 k_{\rm F}$ (all energy scales are sent to 0) 
is, however, in general characterized by a power-law divergence (for repulsive interactions) 
which indicates the breakdown of linear 
response theory. It is a signature of the sensitivity of a LL against single-particle 
perturbations with momentum transfer $2 k_{\rm F}$ (backscattering).\cite{Giamarchi03,Schoenhammer05}  
We note that the character of the density-density correlation function with small momentum 
transfer is not modified by the momentum dependence of the potentials; it is 
given by a $\delta$-function at energy $\omega_c(q)$ instead of $v_c |q|$.    

\subsubsection{Spectra of lattice models}

Directly computing the single-particle spectral function for translational 
invariant microscopic lattice models of 1d correlated fermions still poses a 
formidable challenge of quantum many-body theory. Two promising routes exists. 

The first one is numerical in nature. Using exact diagonalization (ED), 
the (dynamical) density-matrix RG (DMRG), or different types of quantum Monte-Carlo (QMC) approaches, 
valuable information on the spectral function of different models was collected. However, 
the search for power laws requires an exceptional energy resolution. This implies that fairly
large system sizes and low temperatures must be accessible. The model most heavily investigated
is the 1d Hubbard model (and variants of it). Away from half-filling it is known 
to fall into the LL universality class. ED is restricted to systems of a few ten lattice 
sites which leads to a poor energy resolution. Within so-called
cluster perturbation theory, which is ED-based, it was at least possible to observe spin-charge 
separation.\cite{Senechal00} The latter was systematically studied using 
QMC.\cite{Zacher98,Abendschein06} In these studies the finite temperature (and the required analytic continuation) 
turned out to be the main obstacle preventing an analysis of 
$\rho^>(k ,\omega)$ in terms of power-law scaling. Reference \onlinecite{Benthien04} 
contains the first serious attempt in this direction. The authors use (dynamical) DMRG to obtain 
broadened spectra. A scaling analysis as a function of the broadening was interpreted to be consistent 
with power-law behavior of the charge and spin peaks. The quality of the data is, however, not
good enough to either confirm or refute power laws at $k-k_{\rm F} \neq 0$.
Generally speaking the numerical results for lattice models from the LL universality class 
are fully consistent with our conclusions.  

A word of warning is in order. As our analysis of Sect.~\ref{sect_g2_g4} shows
it can be very difficult to undoubtedly confirm or refute power-law behavior of broadened 
finite size data. This holds even though for the TLM we can achieve a comparably high 
energy resolution and obtain data which are sufficiently accurate to employ a very sensitive 
logarithmic derivative.      

A promising analytical approach to the spectral function of 1d models is built upon the special
structure of several of the standard lattice models from the LL universality class, namely the 
existence of an extensive number of local integrals of motion. For this reason several models 
are exactly solvable by Bethe ansatz, which, however, does not imply that the single-particle 
spectral function can (easily) be computed exactly.\cite{Essler05} In a series of papers, 
see Ref.~\onlinecite{Carmelo06} and references therein, $\rho^>(k ,\omega)$ of the 1d Hubbard 
model was investigated using a `pseudo\-fermion dynamical theory' which is based on the Bethe 
ansatz solution. It was reported that the spectral function contains power laws even for 
$k - k_{\rm F} \neq 0$, with exponents which depend on $k - k_{\rm F}$. This finding 
might be related to the `nonlinear Luttinger liquid phenomenology' briefly touched below. 
The spectral function of another exactly solvable 1d model, namely the Calogero-Sutherland 
model showing power laws with momentum dependent exponents was interpreted in the light of 
this phenomenology.\cite{Khodas07} 

We emphasize that our findings do not exclude that specific models from the LL 
universality class might show finite $k - k_{\rm F}$ power laws, however, if so for 
{\it more specific reasons than LL universality.} 
Models with equilibrium dynamics which are restricted  by an extensive 
number of local conserved quantities might be examples for such behavior.  
This is also supported by the observation that 
the  power laws reported on in Refs.~\onlinecite{Benthien04,Carmelo06,Khodas07} 
cannot only be found at  low energies (for very small $|k-k_{\rm F}|$), while 
LL theory is supposed to be applicable in this limit only.

\subsection{Implications for the `nonlinear Luttinger liquid phenomenology'}
\label{sect_con_nonlinLL}

In an extensive series of papers, reviewed in Ref.~\onlinecite{Imambekov12}, 
a novel approach to study corrections to LL behavior by the curvature 
of the single-particle dispersion was developed. It cumulates in the so-called 
`nonlinear Luttinger liquid phenomenology'. The approach is mainly built upon 
an effective field theory which is motivated by lowest order perturbation 
theory. For the single-particle spectral function this phenomenology predicts 
(Fermi edge singularity like) power laws with momentum dependent exponents even at 
$k-k_{\rm F} \neq 0$. However, the field theory relies on the assumption 
of a momentum independent bulk interaction (the momentum dependence of the
interaction with the mobile impurity is considered) and requires ad hoc 
ultraviolet regularization. As we have shown in (linear) Luttinger liquid 
theory the same assumption leads to power-law behavior which is nongeneric 
rather than universal. This raises the question whether the power laws found in 
`nonlinear LL phenomenology' are robust against a curvature of the bulk 
two-particle potential.
  
\subsection{Implications for the interpretation of experimental spectra} 
\label{sect_con_exp}

\begin{figure}[t]
\includegraphics[width=1.\linewidth,clip]{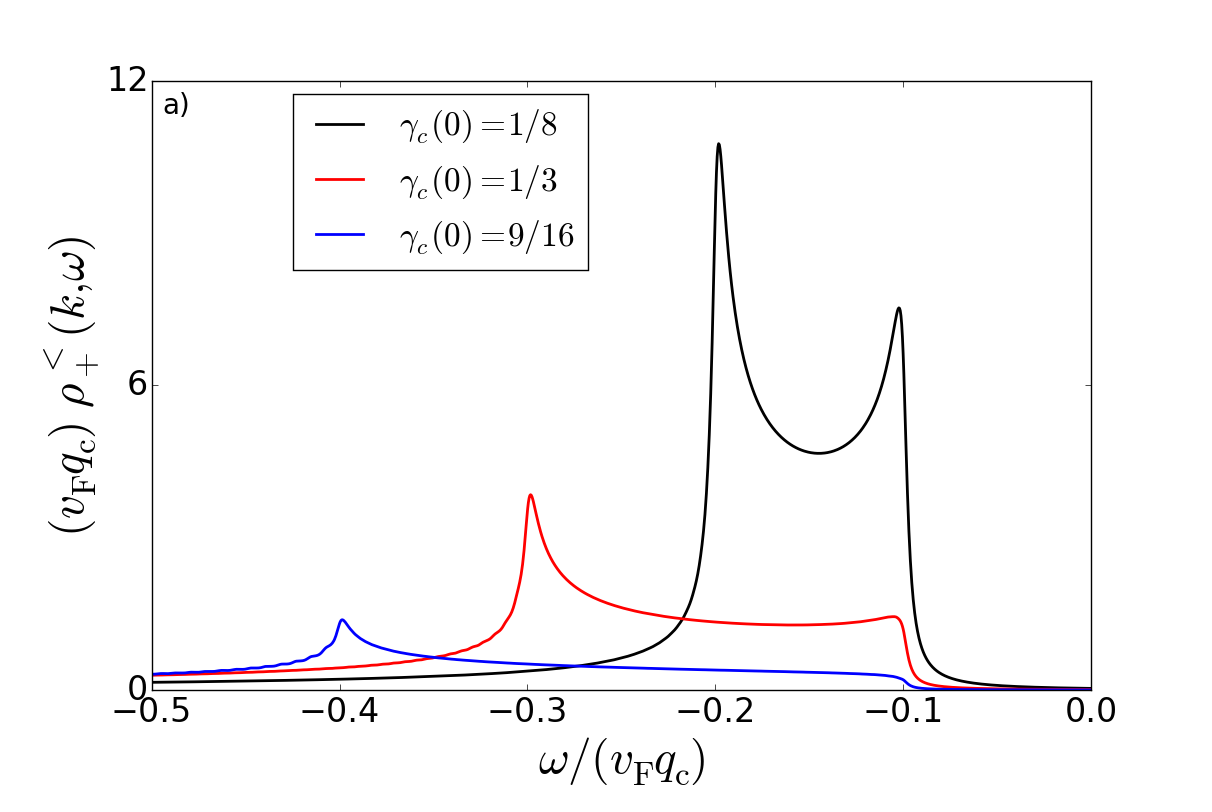}
\includegraphics[width=1.\linewidth,clip]{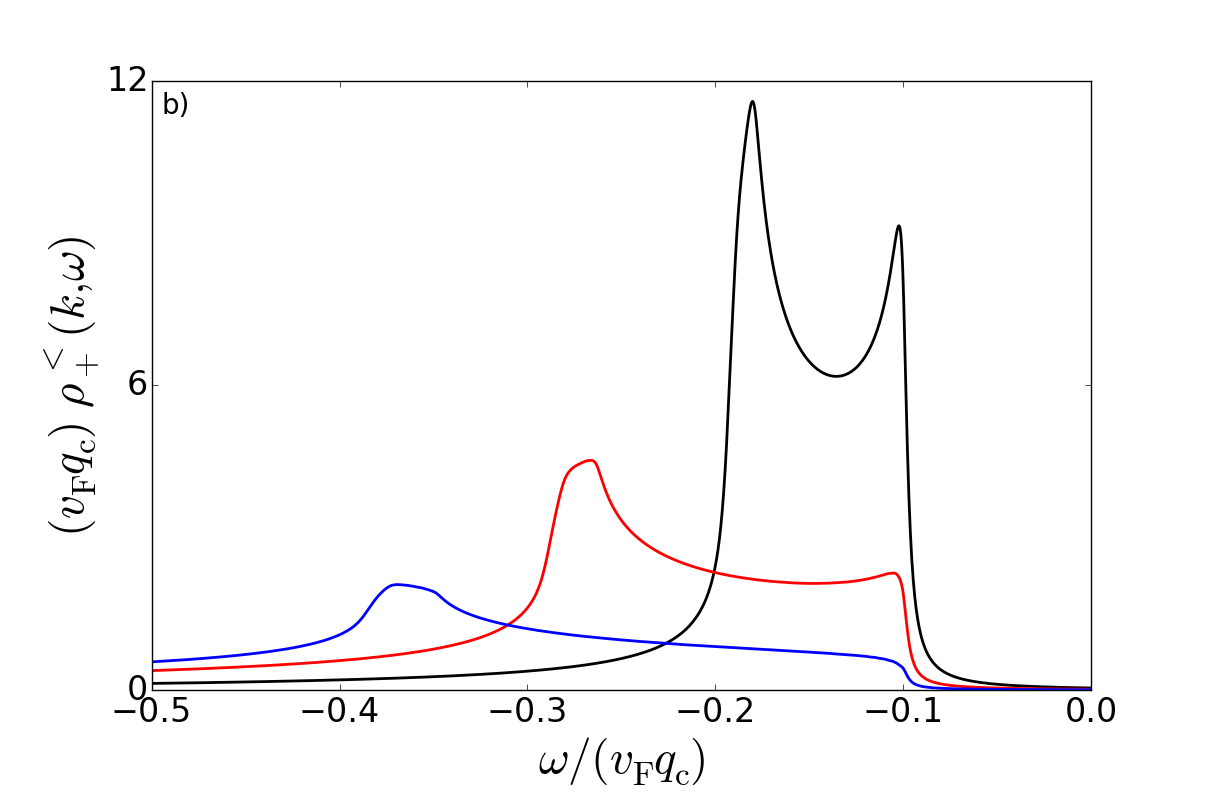}
\caption{(Color online) Photoemission part of the spectral function of 
the spinful TLM at $(k-k_{\rm F})/q_{\rm c}=-0.1$. A spin-independent 
interaction with equal inter- and intra-branch amplitude is 
assumed. The interaction strength is chosen such that $\gamma_c(q=0)$ assumes
the values given in the legend. The other parameters are 
$n_{\rm c} =10^3$ (system size) and $\chi/(v_{\rm F} q_{\rm c})=3 \cdot 10^{-3}$ (broadening). 
In a) the spectra for the Gaussian 
potential are shown, in b) the ones for the exponential 
potential. The small `high-energy' oscillations for larger interactions
in a) are a finite size effect.  
}
\label{fig5}
\end{figure}

When interpreting experimental angular resolved photoemission data on 
quasi-1d metallic materials\cite{Grioni09} the ad hoc regularized spectral function of 
the TLM is often taken paradigmatically. It is e.g. expected that the distribution 
of the spectral weight between the spin and charge peaks as well as the spacing 
between the two must be exactly as in the analytical expressions given in 
Refs.~\onlinecite{Meden92,Voit93,Voit93b,Voit95}. A very recent example 
in which this leads to a putative conflict can be found in Ref.~\onlinecite{Ohtsubo15}. 
Our results for the spinful $g_4$- and the spinless $g_2$-$g_4$-models  
show that the ad hoc regularized spectral function cannot be 
considered as universal. Within the TLM the details of the distribution of spectral 
weight, the peak distance, and the line shape clearly depend on the form of the 
two-particle potential even when considering small $|k-k_{\rm F}|$. Such `details' are 
not part of LL universality and {\it expecting quantitative agreement  
overstresses LL theory.}  We reiterate that all that is known about the $T=0$ spectral 
properties from the latter is that for $\omega \to 0^+$ the scaling relations  
$\rho^>(k_{\rm F},\omega) \sim \omega^{\alpha-1}$ and $\rho^>(\omega) \sim \omega^\alpha$ hold. 
The exponent $\alpha=\gamma_c+\gamma_s$ can be expressed in terms of the LL parameters 
$K_{c/s}$ [see Eq.~(\ref{gammadef}); $K_s =1$ for spin-rotational invariant systems],
which also enter in other `observables'.\cite{Giamarchi03,Schoenhammer05} If they can 
be measured for the same system, consistency checks are possible. LL universality
also makes predictions for the scaling of spectral weight as a function $T$. It
was shown that $\rho^>(\omega=0) \sim T^\alpha$ and $\rho^>(k_{\rm F},\omega=0) 
\sim T^{\alpha-1}$.\cite{Schoenhammer93c,Nakamura97,Orgad01} We note that details 
of the analysis of $\rho_+^>(k,\omega)$ at $T>0$ for the TLM\cite{Nakamura97,Orgad01} 
beyond the above scaling relation should be taken with caution as they rely 
on an ad hoc regularization procedure.

It is tempting to employ the (spinful) TLM spectra computed with proper 
$g_{i,\kappa}(q)$ for comparison to experimental ones beyond the above scaling 
relations. This might lead to a {\it qualitative agreement} of certain features. 
One crucial generic feature we found regardless of the $q$-dependence of the 
two-particle potential are dispersing spin and charge `peaks', however, 
generically not given by power-law singularities. To illustrate this 
we show the spectral function of the spinful TLM for the Gaussian and the 
exponential potential in Figs.~\ref{fig5}a) and b), respectively. This time we 
focus on the {\it photoemission part}  $\rho_+^<(k,\omega)$ for $(k-k_{\rm F})/q_{\rm c} = -0.1$
as this is most easily accessible experimentally. The results were obtained by a 
straightforward generalization
of the recursive procedure put forward for the spinful $g_4$-model 
(see Sect.~\ref{sping4arb}) and the spinless $g_2$-$g_4$-model 
(see Sect.~\ref{nospinarb}).  The system size is set by $n_{\rm c} = 10^3$ and 
the broadening by $\chi/(v_{\rm F} q_{\rm c})=3 \cdot 10^{-3}$. We consider the physically reasonable case 
of a spin-independent interaction with an equal amplitude of the intra- and inter-branch  
parts. This implies $\omega_s(q) = v_{\rm F} q$ and thus $v_s(q=0) = v_{\rm F}$.        
The interaction is varied such that $\gamma_c(q=0)=1/8$, $1/3$, and $9/16$.
Accordingly, $v_c(q=0)$ increases. As for the above discussed limiting cases 
of the spinful $g_4$- and the spinless $g_2$-$g_4$-model, the details of the spectral 
line shape obviously depend on the details of the two-particle potential 
[compare Figs.~\ref{fig5}a) and b)]. The integral over $\rho_+(k,\omega) = 
\rho^>_{+} (k, \omega) + \rho^<_{+} (k, \omega)$ at fixed $k-k_{\rm F}$ must be 
normalized to 1 which implies that for the exponential potential less spectral 
weight is transferred to $ \rho^>_{+} (k, \omega)$ (not shown) than for the Gaussian one. 
For the former the `charge peak' is fairly broad and has an unusal line shape. 
However, the two sets of spectra also share similarities. For increasing interaction 
the spin peak loses weight and deforms into a shoulder like 
structure. To summarize, we expect spin-charge separation to be a robust feature
of quasi-1d metalls which should be observable in photoemission experiments.

\begin{acknowledgements}
We are grateful to Kurt Sch\"onhammer for numerous discussions. This work was supported 
by the DFG via RTG 1995.
\end{acknowledgements}

{}

\end{document}